\documentclass[reprint,amsmath,amssymb,aps,superscriptaddress,nofootinbib,twocolumn,10pt]{revtex4-1}
\usepackage[colorlinks,linkcolor=blue,citecolor=blue,urlcolor=blue]{hyperref}
\usepackage{graphicx,epstopdf}
\usepackage{dcolumn}
\usepackage{verbatim}
\usepackage{bm}
\usepackage{amsmath,amsthm,amsfonts,amssymb,graphicx,multirow,dcolumn,bm,latexsym,soul,nicefrac}
\allowdisplaybreaks[4]
\usepackage{acronym}
\usepackage{enumitem}
\usepackage{array}
\usepackage{multirow}
\usepackage{appendix}
\usepackage{footnote}
\usepackage{amssymb}
\usepackage{subcaption}
\usepackage{calc}
\usepackage{tikz}
\newcommand{\mathD}{\mathrm{D}}
\newcommand{\mathd}{\mathrm{d}}
\newcommand{\assign}{:=}
\newcommand{\nobracket}{}
\newcommand{\nocomma}{}
\newcommand{\mathi}{\mathrm{i}}
\def\nn{\nonumber}

\newacro{GW}{gravitational-wave}
\newacro{GR}{general relativity}
\newacro{PTA}{Pulsar Timing Array}
\newacro{SGWB}{stochastic gravitational-wave background}
\newacro{LIGO}{Laser interferometry Gravitational-Wave Observatory}
\newacro{TDI}{Time Delay Interferometry}
\newacro{TQ}{TianQin}
\newacro{CO}{Compact Object}
\newacro{LISA}{Laser Interferometry Space Antenna}
\newacro{CBC}{compact binary coalescence}
\newacro{BH}{black hole}
\newacro{SBBH}{stellar-mass binary black hole}
\newacro{PN}{post-Newtonian}
\newacro{PSD}{power spectral density}
\newacro{ORF}{overlap reduction function}
\newacro{PLIS}{power-law integrated sensitivity}
\newacro{FIM}{Fisher information matrix}
\newacro{SNR}{signal-to-noise ratio}
\newacro{EMRI}{Extreme Mass Ratio Inspiral}
\newacro{MBH}{Massive Black Hole}
\newacro{LSO}{Last Stable Orbit}
\newacro{SF}{Self-Force}

\newcommand
{\TRC}{MOE Key Laboratory of TianQin Mission,
TianQin Research Center for Gravitational Physics \& School of Physics and Astronomy,
Frontiers Science Center for TianQin,
Gravitational Wave Research Center of CNSA,
Sun Yat-sen University (Zhuhai Campus), Zhuhai 519082, China.}

\begin{document}

\title{Toward Second-Order Self-Force for Eccentric Extreme-Mass Ratio Inspirals in Schwarzschild Spacetime}

\author{Yi-Xiang Wei}
\affiliation{\TRC}
\author{Xian-Long Zhu}
\affiliation{\TRC}
\author{Jian-dong Zhang}
\email{zhangjd9@mail.sysu.edu.cn}
\affiliation{\TRC}
\author{Jianwei Mei}
\affiliation{\TRC}

\begin{abstract}

An \ac{EMRI}, which corresponds to a small compact object inspirals around a massive black hole in the center of a galaxy,
is one of the most important sources for future space-borne \ac{GW} detectors such as TianQin and LISA.
By analyzing the emitted \ac{GW} signals,
we can probe the theory of gravity and the nature of black holes in the strong field region.
To achieve these objectives, the second-order self-force effect should be considered in the waveform modeling.
Up to now, the waveform of \acp{EMRI} including the second-order self-force effect
is only achieved for the circular orbit on Schwarzschild background.
In this work, we generalized the calculation of the second-order self-force
to the eccentric orbits on Schwarzschild spacetime.
We calculated the puncture field, and give the form of two-timescale expansion for the field equations.
The corresponding numerical calculation and programming can be performed based on these results.

\end{abstract}

\maketitle

\section{Introduction}\label{intro}

Multiple space-borne \ac{GW} detectors,
such as TianQin\cite{TianQin:2015yph,Hu:2017yoc,TianQin:2020hid,Luo:2025sos} and LISA\cite{Danzmann:1997hm,LISA:2017pwj,LISA:2024hlh},
are scheduled to begin the operation in the middle 2030s.
These milli-Hertz band detectors are expected to detect a wide variety of \ac{GW} signals
and yield significant accomplishments on fundamental physics, astrophysics, and cosmology
\cite{Luo:2025ewp,Li:2024rnk,LISA:2022kgy,Barausse:2020rsu,LISACosmologyWorkingGroup:2022jok,LISA:2022yao}.
In order to achieve these objectives, an accurate and precise waveform is needed
\cite{LISAConsortiumWaveformWorkingGroup:2023arg} for the sources to avoid possible bias
\cite{Dhani:2024jja,Kapil:2024zdn,Owen:2023mid,Gupta:2024gun} and get a precise measurement on the parameters.

\acp{EMRI} refer to a special class of binary systems with a small mass ratio $\varepsilon=\mu/M$,
that generate gravitational wave signals during the inspiral phase \cite{Finn:2000sy}.
Here $M$ is the mass of the primary, ranging at $10^4 \sim 10^7 M_{\odot}$,
supposed to be a \ac{MBH} in the center of the galaxies.
$\mu$ is the mass of the secondary, a \ac{CO} ranging at $1 \sim 10^2 M_{\odot}$,
could be a neutron star or a stellar mass \ac{BH}.
During the inspiral phase of such a binary system,
the two objects gradually approach each other along a quasi-periodic orbit,
emitting gravitational waves with the amplitude and frequency increasing progressively
until the system's orbit evolves to the \ac{LSO}\cite{Barack:2018yvs}.

For typical \acp{EMRI}, the frequency of the radiated gravitational waves will fall within the millihertz band,
making them one of the primary targets for space-borne \ac{GW} detectors.
In a detectable \ac{EMRI} signal,
it would contain about $10^6$ cycles for the \ac{CO} moving in the strong field region,
before the \ac{CO} plunges into the central \ac{MBH}.
Different astronomical models are expected to have different detection rates for \ac{EMRI},
ranging from nearly zero to tens of thousands per year \cite{Babak:2017tow,Fan:2020zhy}.

\ac{EMRI} systems have unique significance in astronomical observations.
Due to the long inspiral phase, the \acp{CO} will have highly complex orbits
which could reveal numerous details of
the strong gravitational field \cite{Tan:2024utr,Zi:2022hcc,Zi:2021pdp},
the astrophysical properties of the MBH \cite{Li:2025zgo,Cui:2025bgu},
and the surrounding environment \cite{Zhang:2024ugv,Lyu:2024gnk}.
It's expected to have a very high precision on the measurement of the corresponding parameters \cite{Babak:2017tow,Fan:2020zhy},
which also means a very high requirement on the accuracy of the waveforms and data analyzing methods.
However, due to the very long and complex orbital motion,
to generate the waveforms for \acp{EMRI} fast and accurate is still a big challenge.

Under the limit of $\varepsilon \rightarrow 0$, The secondary behaves as a test particle,
moving along the geodesics on the background spacetime of the primary.
In this paper we consider the effect at the linear or higher order of $\varepsilon$,
the orbit will deviate from geodesic due to the emission of the \acp{GW}.
If we expand the metric $\mathfrak g_{\mu \nu}$ in terms of $\varepsilon$ based on the background $g_{\mu \nu}$,
we will have
\begin{equation*}
  \mathfrak g_{\mu \nu} = g_{\mu \nu} + \varepsilon h_{\mu \nu}^{(1)}
  + \varepsilon^2 h_{\mu \nu}^{(2)} +\mathcal{O} (\varepsilon^3)
\end{equation*}
Thus the geodesic equation for will become
\begin{equation*}
  \frac{\mathD^2 z^{\mu}}{\mathd \tau^2}  =  \varepsilon f^{(1) \mu} +
  \varepsilon^2 f^{(2) \mu} +\mathcal{O} (\varepsilon^3),
\end{equation*}
where the covariant derivative is defined in terms of background metric $g_{\mu\nu}$,
and $f^{(n) \mu}$ are the \ac{SF} terms as functions of perturbations $h_{\mu\nu}$ and trajectory $z^{\mu}$.

The analytical expression for the 1SF term was first derived by
\cite{Mino:1996nk,Quinn:1996am} independently in 1996,
and the corresponding equation of motion for the secondary on the background is known as the MiSaTaQuWa equation.
The analytical expression for 2SF term was given by \cite{Pound:2012nt,Gralla:2012db} in 2012.
A general algorithm for the analytical calculation of \ac{SF} to arbitrary orders is provided in \cite{Pound:2015tma}.
A rough estimation shows that 2\ac{SF} should be considered in the waveform modeling,
if we want the accuracy on the phase to achieve phase
errors of $\mathcal O (\varepsilon)$ for matched filtering.
The numerical result \cite{Burke:2023lno} shows that the 2nd \ac{SF} is enough for data analysis.

By considering the behavior under time reversal,
we can separate the n-th order \ac{SF} $f^{(n) \mu}$ into
a antisymmetric dissipative part $f_{\mathrm{diss}}^{(n) \mu}$
and a symmetric conservative part $f_{\mathrm{cons}}^{(n) \mu}$ as
\begin{equation*}
  f^{(n) \mu}  = f_{\mathrm{diss}}^{(n) \mu} + f_{\mathrm{cons}}^{(n) \mu},
\end{equation*}
The effect of $f_{\mathrm{diss}}^{(n) \mu}$ will accumulate during the evolution,
and will be the same order with with $f_{\mathrm{cons}}^{(n - 1) \mu}$.
The analysis in \cite{Hinderer:2008dm} shows that
only the dissipative part of 2nd order \ac{SF} need to be considered in the waveform modeling.
If we expand the accumulated phase as
\begin{equation*}
  \varphi  =  \frac{1}{\varepsilon} \varphi^{(- 1)} + \varphi^{(0)} +
  \varepsilon \varphi^{(1)} +\mathcal{O} (\varepsilon^2),
\end{equation*}
only $f_{\mathrm{diss}}^{(1) \mu}$ will contribute to the adiabatic (0PA) term $\varphi^{(- 1)}$,
while both $f_{\mathrm{cons}}^{(1) \mu}$ and $f_{\mathrm{diss}}^{(2) \mu}$ will contribute to
the first-order post-adiabatic (1PA) term $\varphi^{(0)}$.

The 1st order \ac{SF} waveform include $f_{\mathrm{diss}}^{(1) \mu}$ and $f_{\mathrm{cons}}^{(1)\mu}$,
considered the effect of resonant orbits for generic Kerr orbits
using fast algorithm is recently given \cite{Lynch:2024ohd}.
However, the 1PA waveform including $f_\mathrm{diss}^{(2) \mu}$
is still in its early stages due to the complexity of the calculations.
In the calculation of $f_\mathrm{diss}^{(2) \mu}$,
the puncture field method is used to obtain the regular part of the metric perturbation \cite{Warburton:2013lea,Wardell:2015ada,Miller:2023ers},
while the two timescale expansion is used to decouple the oscillation and dissipation effect,
thereby accelerating the numerical calculation.
The two timescale method was first introduced into the study of \acp{EMRI} in \cite{Hinderer:2008dm}.
A general scheme to generate \ac{EMRI} waveforms at 1PA order with two timescale methods
has been presented in \cite{Pound:2021qin} for general bounded orbits on Kerr background.
However, due to the complexity of the specific calculations,
the 1PA waveform has only been achieved for quasicircular orbits on Schwarzschild spacetime
\cite{Miller:2020bft,Warburton:2021kwk,Wardell:2021fyy,Warburton:2024xnr}.
Since most real EMRIs have a large orbital eccentricity\cite{Amaro-Seoane:2012jcd},
there is a strong demand to push those calculations into eccentric orbits.

In this work, we generalize the calculation of the 2nd order \ac{SF}
to eccentric orbit for \acp{EMRI} around a Schwarzschild blackhole.
By employing the osculating geodesics method,
the evolution of coordinates were transformed into
the evolution of the parameters of Schwarzschild geodesics.
We further introduce the action angle variables to decouple
the fast orbital motion from the slow radiation-driven inspiral,
make it possible to simplify the equations into a two timescale expansion form,
which will significantly reduce the computational complexity.
We also derived the two timescale expansion form of the field equation,
and calculated the puncture field for eccentric orbit.

This paper is orgnized as follows.
We begin with a review on the eccentric orbits on Schwarzschild spacetime in Sec \ref{geodesic}.
The two-timescale form of the field equation for eccentric orbits is obtained in Sec \ref{twoscalefield}.
The puncture fields for eccentric orbits are given in Sec \ref{puncture}.
The two-timescale expansion of the orbital equation of motion
for eccentric orbits is calculated in Sec \ref{twoscalefield}.
Finally we give some comments in Sec \ref{con}.

Throughout the article, We set $G = c = 1$ for geometrized unit,
and use a positive Lorentz signature $(-, +, +, +)$ for the metric.

\section{Eccentric Geodesic for Schwarzschild Black Hole}\label{geodesic}

Since the geodesic is regarded as the zeroth-order approximation for \acp{EMRI},
We first give some properties and definitions for bounded eccentric geodesics
on Schwarzschild spacetime
\begin{eqnarray}
 \mathd s^2& =& - \left( 1 - \frac{2 M}{r} \right) \mathd t^2 + \left( 1 -
   \frac{2 M}{r} \right)^{- 1} \mathd r^2 +\nn\\
   && r^2 (\sin^2 \theta \mathd
   \varphi^2 + \mathd \theta^2).
\end{eqnarray}
The coordinates of a test particle are denoted as
$x_{\text{p}}^{\mu} (\tau) =\left\{ t_{\text{p}} (\tau),
r_{\text{p}} (\tau), \theta_{\text{p}} (\tau),
\phi_{\text{p}} (\tau) \right\}$,
where $\tau$ is the proper time for the test particle.
The geodesic on Schwarzschild has been well studied in previous works such as \cite{darwin1959gravity,darwin1961gravity,Cutler:1994pb},
and the introduction here aims to provide the definitions of
some quantities used in the following calculations.

Due to the spherical symmetry of Schwarzschild spacetime, we can set
$\theta_{\text{p}} (\tau) \equiv \frac{\pi}{2}$ without loss of generality.
Thus the geodesic equation can be written as
\begin{eqnarray}
  \frac{\mathd t_{\text{p}}}{\mathd \tau} & = & E \left( 1 - \frac{2M}{r_{\text{p}}} \right)^{- 1} \\
  \left( \frac{\mathd r_{\text{p}}}{\mathd \tau} \right)^2 & = & E^2 -
  \left( 1 - \frac{2 M}{r_{\text{p}}} \right)  \left( \frac{J^2}{r_{\text{p}}^2} + 1 \right) \\
  \frac{\mathd \phi_{\text{p}}}{\mathd \tau} & = & \frac{J}{r_{\text{p}}^2},
\end{eqnarray}
where the $E, J$ are the energy and angular momentum for the test particle.

Since $\mathd r_{\text{p}}/\mathd \tau$ will change its sign
for every half cycle as it approached to the apsides,
a better way to describe the radial motion is to
use the relativistic anomaly $\chi$ instead of $\tau$:
\begin{equation}
  r_{\text{p}} (\chi) = \frac{pM}{1 + e \cos \chi},
\end{equation}
where
\begin{align}
  p = \frac{2 r_{\max} r_{\min}}{(r_{\max} + r_{\min}) M} & , & e =
  \frac{r_{\max} - r_{\min}}{r_{\max} + r_{\min}}
\end{align}
are the dimensionless semi-latus rectum and eccentricity of the orbit. $r_{\max}$ and $r_{\min}$ are the apoapsis and periapsis respectively.
Then the constant of motion $E$ and $J$ can be represened as
\begin{equation}
  J = \frac{p M}{\sqrt{p - 3 - e^2}},~~~~~~
  E = \sqrt{\frac{(p - 2)^2 - 4 e^2}{p (p - 3 - e^2)}}
\end{equation}
Then the geodesic equations are
\begin{eqnarray}
  \frac{\mathd t_{\text{p}}}{\mathd \chi} & = & \frac{r_{\text{p}}^2}{M [p - 2 - 2 e \cos
  \chi]}  \sqrt{\frac{(p - 2)^2 - 4 e^2}{p - 6 - 2 e \cos \chi}} \label{Ellp}\\
  \frac{\mathd \phi_{\text{p}}}{\mathd \chi} & = & \sqrt{\frac{p}{p - 6 - 2 e
  \cos \chi}}\label{Ellf}
\end{eqnarray}
These equations can be integrated directly to obtain the analytical solutions with elliptic integrals.
\begin{eqnarray}
  \phi_{\text{p}} (\chi) & = & 2 \sqrt{\frac{p}{p - 6 + 2 e}} \left[ K
  \left( \frac{4 e}{p - 6 + 2 e} \right) \right. \nn\\
  && -\left. F \left( \frac{\pi}{2} -
  \frac{\chi}{2} | \frac{4 e}{p - 6 + 2 e} \right) \right];\\
  t_{\text{p}} (\chi) & = & I_2 \left( e, \frac{2 e}{p - 2}, \frac{2 e}{p - 6}
  ; \chi \right).
\end{eqnarray}
The definitions of the elliptic integrals $K,F,I_2$ are given in Appendix A.

The 4-velocity of particle is therefore functions of $(p,e,\chi)$:
\begin{eqnarray}
  u^r (p, e, \chi) & = & e \sin \chi \sqrt{\frac{p - 6 - 2 e \cos \chi}{p (p -
  3 - e^2)}} ;\label{ur}\\
  u^t (p, e, \chi) & = & \frac{p}{p - 2 - 2 e \cos \chi} \sqrt{\frac{(p - 2)^2
  - 4 e^2}{p (p - 3 - e^2)}} ;\label{ut}\\
  u^{\varphi} (p, e, \chi) & = & \frac{(1 + e \cos \chi)^2}{p M}
  \sqrt{\frac{p}{p - 3 - e^2}} .\label{uphi}
\end{eqnarray}

The period of radial libration corresponds to $\Delta \chi = 2 \pi$,
and we can define
\begin{equation}
T_r = t_{\text{p}}  (2 \pi),
\end{equation}
During the radial period, the growth of azimuthal angle is
\begin{equation}
\Phi = \phi_{\text{p}}  (2 \pi) = 4 \sqrt{\frac{p}{p - 6 + 2 e}} K
  \left( \frac{4 e}{p - 6 + 2 e} \right).
\end{equation}
We can further define the associated fundamental frequencies
\begin{equation}
  \Omega_r  =  \frac{2 \pi}{T_r},~~~
  \Omega_{\phi}  =  \frac{\Phi}{T_r} = \frac{\Phi}{2 \pi} \Omega_r .
\end{equation}
To seperate the effects due to dissipative forces in long-timesacle
and those from conservative forces on short-timescale periodic motion,
following \cite{Hinderer:2008dm},
we define $\tilde{t} = \varepsilon t$ as the long time variable,
while the short time variables are the action-angles
\begin{equation}
  q_t  =  t,~~~
  q_r  = \Omega_r t,~~~
  q_{\phi}  =  \Omega_{\phi} t
\end{equation}

\section{The Two Time Expansion of field equations}\label{twoscalefield}

Two timescale exansion is a technique commonly encountered in nonlinear equation analysis,
primarily applicable to periodic systems subject to dissipative perturbations\cite{kevorkian2012multiple}.
There are typically two distinct characteristic time scales:
the short time scale of the periodic oscillations with almost fixed parameters
and the long time scale over which dissipative effects change parameters of the system.
Usually the dissipative effects are perturbations of order $\varepsilon$,
then the latter time scale is generally at the order of $\sim 1 / \varepsilon$.
To capture these two time scales, we can introduce new variables
\begin{equation}
  \tau = [1 +\mathcal{O}(\varepsilon)] t,~~~\tilde{t} = \varepsilon t
\end{equation}
to represent the short and long time scale, respectively.
So the functions of $t$ can be rewritten as the functions of $(\tau, \tilde{t})$,
and then be expanded in terms of $\varepsilon$, which is known as two timescale expansion.

Typically, this expansion yields solutions of the form
\begin{eqnarray}
  f (\tau, \tilde{t}) & = & A (\tilde{t}) \cos (\tau + B (\tilde{t})),
  \nonumber
\end{eqnarray}
This form clearly shows the influence of the two different time scales on the system:
the shorter time scale $\tau$ contributes to the rapid oscillations,
while the amplitude $A (\tilde{t})$ and phase $B (\tilde{t})$ will
vary with the dissipative effects over the longer time scale $\tilde{t}$.
Thus the effects of rapid oscillations and long-term dissipative evolution are decoupled.
This will lead to a individually analysis of the oscillations in the frequency domain,
thereby accelerating the computational process.

\subsection{The perturbative expansion}
Since $\epsilon$ is small,
we can expand the Einstein equations to perturbative equations at each order of $\epsilon$.
Then the Einstein tensor $G_{\mu \nu}  [g + h]$ can be expanded as
\begin{equation}
  G_{\mu \nu}  [g+h]  =  G_{\mu \nu} [g] + \varepsilon G_{\mu
  \nu}^{(1)} [h] + \varepsilon^2 G_{\mu \nu}^{(2)} [h] +\mathcal{O}
  (\varepsilon^3)
\end{equation}
with
\begin{eqnarray}
  G_{\mu \nu}^{(1)} [h] & = & \left( \delta_{\mu}^{\alpha}
  \delta_{\nu}^{\beta} - \frac{1}{2} g_{\mu \nu} g^{\alpha \beta} \right)
  R_{\alpha \beta}^{(1)} [h] \\
  G_{\mu \nu}^{(2)} [h] & = & \left( \delta_{\mu}^{\alpha}
  \delta_{\nu}^{\beta} - \frac{1}{2} g_{\mu \nu} g^{\alpha \beta} \right)
  R_{\alpha \beta}^{(2)}  [h] \nn\\
  &  & - \frac{1}{2}  (h_{\mu \nu} g^{\alpha \beta} - g_{\mu \nu} h^{\alpha
  \beta}) R_{\alpha \beta}^{(1)} [h]
\end{eqnarray}
while the linear term $R_{\alpha \beta}^{(1)} [h]$ and
the quadratic term $R_{\alpha \beta}^{(2)} [h]$ have the form of
\begin{eqnarray}
  R_{\alpha \beta}^{(1)} [h] & = & - \frac{1}{2}  \left( \Box h_{\alpha \beta}
  + 2 R_{\alpha \phantom{\mu} \beta \phantom{\nu}}^{\phantom{\alpha} \mu
  \phantom{\beta} \nu} h_{\mu \nu} - 2 \nabla^{\mu} \nabla_{(\beta}
  \bar{h}_{\alpha) \mu} \right) \nn\\
  R_{\alpha \beta}^{(2)}  [h] & = & \frac{1}{4} \nabla_{\alpha} h^{\mu \nu}
  \nabla_{\beta} h_{\mu \nu} + \frac{1}{2} \nabla^{\nu} h_{\phantom{\mu}
  \beta}^{\mu} \nabla_{[\nu} h_{\mu] \alpha} \nn\\
  &  & - \frac{1}{2} \nabla_{\nu}  \bar{h}^{\mu \nu}  (2 \nabla_{(\beta}
  h_{\alpha) \mu} - \nabla_{\mu} h_{\alpha \beta}) \nn\\
  &  & - \frac{1}{2} h^{\mu \nu}  (\nabla_{\nu} \nabla_{(\beta} h_{\alpha)
  \mu} - \nabla_{\mu} \nabla_{\nu} h_{\alpha \beta}\nonumber\\
  && - \nabla_{\alpha}  \nabla_{\beta} h_{\mu \nu})
\end{eqnarray}
Then, according to the Einstein's equation, we will have
\begin{eqnarray}
  G_{\mu \nu} [g] & = & 0 \nn\\
  G_{\mu \nu}^{(1)} [h^{(1)}] & = & 8 \pi T^{(1)}_{\mu \nu} \nn\\
  G_{\mu \nu}^{(1)} [h^{(2)}] & = & 8 \pi T^{(2)}_{\mu \nu} - G_{\mu
  \nu}^{(2)} [h^{(1)}]
\end{eqnarray}

The perturbation of the metric $h_{\mu \nu}$ can be separated into two parts,
\begin{equation}
  h_{\mu \nu} = h_{\mu \nu}^{\text{R}} + h_{\mu \nu}^{\text{S}},
\end{equation}
where the singular part $h_{\mu \nu}^{\text{S}}$ encodes the structure information of the small object,
while the regular part $h_{\mu\nu}^{\text{R}}$ encodes the \ac{GW} and effect the self-force.

To give the expansion of $T_{\mu \nu}$, We can introduce a effective metric
\begin{equation}
  \breve{g}_{\mu \nu}  =  g_{\mu \nu} + \varepsilon h^{\text{R}}_{\mu \nu}
\end{equation}
and the geodesic equation can be expressed in the effective metric as
\begin{equation}
  \frac{\breve{\mathD}^2 z^{\mu}}{\mathd \breve{\tau}^2}  =  0
\end{equation}
Along the geodesic, the integration
\begin{equation}
  \breve{\tau}  =  \int_{\gamma} \sqrt{- \breve{g}_{\rho \sigma}
  \frac{\mathd z^{\rho}}{\mathd \tau}  \frac{\mathd z^{\sigma}}{\mathd \tau}}
  \mathd \tau = \int_{\gamma} \sqrt{- \breve{g}_{\rho \sigma} u^{\rho}
  u^{\sigma}} \mathd \tau,
\end{equation}
will lead to
\begin{eqnarray}
  \frac{\mathd \tau}{\mathd \breve{\tau}} & = & \frac{1}{\sqrt{- \left[
  g_{\rho \sigma} + \varepsilon h_{\rho \sigma}^{\text{R}} \right] u^{\rho}
  u^{\sigma}}}\nn\\
  & = & \frac{1}{\sqrt{1 - \varepsilon h_{\rho \sigma}^{\text{R}} u^{\rho}
  u^{\sigma}}}\nn\\
  & \approx & 1 + \frac{1}{2} \varepsilon u^{\rho} u^{\sigma} h_{\rho
  \sigma}^{\text{R}} +\mathcal{O} (\varepsilon^2)
\end{eqnarray}

It has been shown in \cite{Detweiler:2011tt} that the stress-energy of the secondary
in the effective metric can be treat as a point-like particle as
\begin{eqnarray}
  T_{\mu \nu} & = & \varepsilon M \int_{\gamma} \breve{u}_{\mu}
  \breve{u}_{\nu}  \breve{\delta}  [x, z (\breve{\tau})] \mathd \breve{\tau}\\
  & = & \varepsilon M \int_{\gamma} \breve{g}_{\mu \alpha}  \breve{g}_{\nu
  \beta} u^{\alpha} u^{\beta}  \frac{\delta^{(4)}  [x - x_p
  (\breve{\tau})]}{\sqrt{- \breve{g}}} \left( \frac{\mathd \tau}{\mathd
  \breve{\tau}} \right) \mathd \tau \nn
\end{eqnarray}
thus we have
\begin{eqnarray}
  T^{(1)}_{\mu \nu} & = & M \int_{\gamma} \frac{\delta^{(4)}  [x - x_p
  (t)]}{\sqrt{- g}}  \frac{u^{\alpha} u^{\beta}}{u^t} \mathd t\\
  T^{(2)}_{\mu \nu} & = & M \int_{\gamma} \frac{\delta^{(4)}  [x - x_p
  (t)]}{u^t  \sqrt{- g}} \times\nn\\
  & & \left[ \frac{1}{2} u_{\mu} u_{\nu} (u^{\rho}
  u^{\sigma} - g^{\rho \sigma}) h_{\rho \sigma}^{\text{R(1)}} \right.\nn\\
  &&\left.+ u^{\alpha}
  u_{\nu} h^{\text{R(1)}}_{\mu \alpha} + u^{\alpha} u_{\mu} h^{\text{R}
  (1)}_{\nu \alpha} \right] \mathd t
\end{eqnarray}

\subsection{The two timescale expansion}

Here we follow the notions in \cite{Miller:2023ers}.
The frequency domain analysis for eccentric orbits involves two distinct fundamental frequencies:
$\Omega_r$ and $\Omega_{\phi}$ .
And both frequencies evaluate slowly at the timescale of $\tilde{t}$.
Then the frequencies of the modes for eccentric motion will be\cite{Cutler:1994pb}
\begin{equation}
  \omega_{m, k}  =  m \Omega_{\phi} + k \Omega_r .
\end{equation}
Following the definitions in \cite{Miller:2023ers}, we choose the hyperboloidal time variable $s$ as
\begin{equation}
  s (t, r)  =  t - k (r^{\ast}),
\end{equation}
where $r^{\ast} = r + 2 M \ln \left( \frac{r}{2 M} - 1 \right)$ is the tortise time
and $k$ is the ``height function'' for which satisfied the conditions
\begin{eqnarray}
  k (r^{\ast}) \rightarrow + r^{\ast} & {for} & r^{\ast} \rightarrow +\infty\\
  k (r^{\ast}) \rightarrow - r^{\ast} & {for} & r^{\ast} \rightarrow - \infty
\end{eqnarray}
The height function is to ensure that the slices of constant $s$ are hyperboloidal
at null infinity and horizon.
Then the evolution of action-angles are
\begin{equation}
      \frac{\mathd q_{\phi}}{\mathd s}=\Omega_{\phi} (\mathcal{J}^I), ~~~~~\frac{\mathd q_r}{\mathd s}=\Omega_r (\mathcal{J}^I), \\
\end{equation}
while the evolution of the orbital frequencies are
\begin{eqnarray}
  \frac{\mathd}{\mathd s} \Omega_{\phi} (\mathcal{J}^I) = \varepsilon
  F_{\phi}^{(0)}  (\Omega_{\phi}, \Omega_r) + \varepsilon^2 F_{\phi}^{(1)}
  \left( \Omega_{\phi}, \Omega_r {, \mathcal{J}^I} \right)\\
  \frac{\mathd}{\mathd s} \Omega_r (\mathcal{J}^I) = \varepsilon F_r^{(0)}
  (\Omega_{\phi}, \Omega_r) + \varepsilon^2 F_r^{(1)}  \left( \Omega_{\phi},
  \Omega_r {, \mathcal{J}^I} \right)
\end{eqnarray}
 $\mathcal{J}^I =(\Omega_r, \Omega_{\phi}, p, e, \varepsilon \delta M, \varepsilon \delta J)$
are the parameters evolve at long timescale,
including the parameters of the geodesic $(p, e)$,
the orbital frequencies $(\Omega_{\phi}, \Omega_r)$,
and the corrections on the parameters for the background metric
$(\varepsilon \delta M, \varepsilon\delta J)$ at 1PA order.

The message about the inspiral phase is contained within the two action-angles $q_A= (q_r, q_{\phi})$ ,
which characterize the fast evolution.
$\mathcal{J}^I$ are functions of $s$ and will be constants for fixed $s$.
Then we can expand tensors with fixed $s$ as
\begin{eqnarray}
  &&h_{\mu \nu}  =  \varepsilon h_{\mu \nu}^{(1)}  (x^i, q_A, \mathcal{J}^I) +\mathcal{O}
  (\varepsilon^2),  \label{hex}\\
  && T_{\mu \nu}  =  \varepsilon T_{\mu \nu}^{(1)} (x^i, q_A, \mathcal{J}^I) +\mathcal{O}
  (\varepsilon^2), \label{Tex}
\end{eqnarray}
where $x^i = (r, \theta, \phi)$ is the spatial coordinates
and all functions on the right hand are functions of $(q_r, q_{\phi})$ with a period of $2\pi$.
Then, we can get the Fourier expansions as
\begin{eqnarray}
  h_{\mu \nu}^{(n)} & = & \sum_{m, k = - \infty}^{+ \infty} h_{\mu \nu}^{(n,
  m, k)}  (x^i, \mathcal{J}^I) e^{- \mathi (mq_{\phi} + kq_r)}
  \label{hf}\\
  T_{\mu \nu}^{(n)} & = & \sum_{m, k = - \infty}^{+ \infty} T_{\mu \nu}^{(n,
  m, k)} (x^i, \mathcal{J}^I) e^{- \mathi (mq_{\phi} + kq_r)} \label{Tf}
\end{eqnarray}
By transform the partial derivatives of the coordinates into
the partial derivatives of $\mathcal{J}^I$ and $q_A$ as
\begin{equation}
  \partial_{\alpha}  =  \delta_{\alpha}^i \partial_i + (\partial_{\alpha} s)
  \left( \Omega_{\phi}  \frac{\partial}{\partial q_{\phi}} + \Omega_r
  \frac{\partial}{\partial q_r} + \frac{\mathd \mathcal{J}^I}{\mathd s}
  \frac{\partial}{\partial \mathcal{J}^I} \right),
\end{equation}
we can expand $G_{\mu\nu}^{(n)}$ in terms of $\varepsilon$,
and thus denote the coefficients of $\varepsilon^j$ as the \ $G_{\mu \nu}^{(n, j)}$.
Then we will have
\begin{eqnarray}
  G_{\mu \nu}^{(1, 0)} [h^{(1)}] & = & 8 \pi T^{(1)}_{\mu \nu}  \label{G10}\\
  G_{\mu \nu}^{(1, 0)} [h^{(2)}] & = & 8 \pi T^{(2)}_{\mu \nu} - G_{\mu
  \nu}^{(2, 0)} [h^{(1)}, h^{(1)}]\nn\\
  && - G_{\mu \nu}^{(1, 1)} [h^{(1)}]
  \label{G102}
\end{eqnarray}

Using the Lorenz gauge
\begin{equation}
  \nabla^{\alpha}  \bar{h}_{\alpha \mu} = 0,
  \label{Z0}
\end{equation}
we can get the relaxed Einstein equations.
Then we can use the Fourier expansions \eqref{hf} and \eqref{Tf}
to obtain the equations of $h_{\mu \nu}^{(n, m, k)}$,
which corresponding to the replacement
\begin{eqnarray}
  \Omega_{\phi}  \frac{\partial}{\partial q_{\phi}} + \Omega_r
  \frac{\partial}{\partial q_r} & \rightarrow & - \mathi (m \Omega_{\phi} + k
  \Omega_r) = - \mathi \omega_{m, k}
\end{eqnarray}
for every mode of \eqref{G10} and \eqref{G102}.

So, for a fixed $s$, we will have a series differential equations
for $h_{\mu \nu}^{(n, m, k)} (x^i ;\mathcal{J}^I)$,
and the equations can be solved numerically.
The detailed form can be found in \cite{Miller:2023ers}.
Then we can obtain the metric perturbations $h_{\mu \nu}^{(n)}$,
and use it to calculate the \ac{SF}.

\section{puncture field for eccentric orbits}\label{puncture}

As the argument in \cite{Mathews:2021rod}, for the calculation at 1PA order,
the only structure information of the secondary which will influence the waveform are the mass and spin.
In this work, we will only consider the mass for brevity.
Then the secondary can be treated as a point mass,
and $h_{\mu \nu}$ will diverge as it approaches to the secondary.

To avoid the divergence in numerical calculations,
a mode-sum method was introduced by regularize each mode independently\cite{Barack:2001ph}.
However, due to the stronger divergence of the 2nd \ac{SF},
this method is inapplicable for the 1PA calculations under Lorenz guage.
Recent studies propose a highly regular gauge choice to alleviate the
divergence and enable the mode-sum regularization\cite{Upton:2021oxf},
but this approach has not yet been put into practice.

Currently, the most widely used method is the puncture field proposed in \cite{Barack:2007jh,Vega:2007mc}.
To avoid the numerical divergence, a puncture field $h_{\mu\nu}^{\mathcal{P}}$ is introduced.
The puncture field should behave like the $h_{\mu \nu}^{\text{S}}$ as
it approaches to the worldline of secondary $\gamma$:
\begin{eqnarray}
  \lim_{x \rightarrow \gamma} \left( h_{\mu \nu}^{\mathcal{P}} - h_{\mu
  \nu}^{\text{S}} \right) & = & 0 \label{hp0} \\
  \lim_{x \rightarrow \gamma} \left( \nabla_{\alpha} h_{\mu \nu}^{\mathcal{P}}
  - \nabla_{\alpha} h_{\mu \nu}^{\text{S}} \right) & = & 0 \label{hp1}
\end{eqnarray}
Then we can define the residual field as
\begin{equation}
  h_{\mu \nu}^{\mathcal{R}} = h_{\mu \nu} - h_{\mu \nu}^{\mathcal{P}}
\end{equation}
and separate the $h_{\mu \nu}$ at each order:
\begin{equation}
  h^{(n)}_{\mu \nu}  =  h_{\mu \nu}^{\mathcal{P} (n)} + h_{\mu
  \nu}^{\mathcal{R} (n)}
\end{equation}
Thus we can use $h_{\mu \nu}^{\mathcal{R}}$ instead of $h_{\mu \nu}^{\text{R}}$
to calculate the \ac{SF} \cite{Pound:2012nt}.

The practical covariant puncture formulas in Lorenz gauge for 2nd SF are given in \cite{Pound:2014xva},
here we follow their procedure to give a concentre coordinate form for eccentric Schwarzschild orbits.
Recently, a covariant formulas in the highly regular gauge is given in\cite{Upton:2023tcv}.
However, since the field equations are expanding in Lorenz gauge,
we will follow \cite{Pound:2014xva} for simplicity.

\subsection{The Construction of Effective Source}

To obtain the analytic form of the puncture field $h_{\mu \nu}^{\mathcal{P}}$,
one can expand the singular field near the worldline
and truncate it at the order of $\lambda$, which is the distance parameter.
Since the decomposition of the perturbative field into the singular and regular part is artificial,
there could be many different kinds of choices to define a singular field.
A special choice proposed in \cite{Detweiler:2002mi} is suitable for \ac{SF} calculations,
where the regular field satisfies a homogeneous wave equation.
We will follow the notions in \cite{Poisson:2011nh} to calculate the singular field.

The trace reversal singular field $\bar{h}_{\mu\nu}^{\text{S}}$ has a covariant expansion,
\begin{equation}
  \bar{h}_{\mu \nu}^{\text{S}}  =  4 \mu g^{\phantom{\mu} \bar{\alpha}}_{\mu}
  g^{\phantom{\nu} \bar{\beta}}_{\nu}  \left[ \frac{u_{\bar{\alpha}}
  u_{\bar{\beta}}}{\bar{s}} +\mathcal{O}(\lambda) \right],
\end{equation}
where the bar over index denotes the point on the worldline and $g^{~\bar{\alpha}}_\mu(x,\bar{x})$
is the parallel transport bi-vector\cite{Poisson:2011nh}.
$u_{\bar{\alpha}}$ is the 4-velocity of the particle at $\bar{x}$ and
\begin{equation}
  \bar{s}^2 = (g_{\bar{\alpha}  \bar{\beta}} + u_{\bar{\alpha}}
  u_{\bar{\beta}}) \sigma^{\bar{\alpha}} \sigma^{\bar{\beta}}
\end{equation}
where $\sigma (x, \bar{x})$ is the Synge world function and
$\sigma_{\bar{\alpha}} = \partial_{\bar{\alpha}} \sigma$ as defined in\cite{Poisson:2011nh}.
An exact form of the $\bar{h}_{\mu \nu}^{\text{S}}$
up to $\mathcal{O} (\lambda^4)$ is given in \cite{Heffernan:2012su}.

For Schwarzschild background, we need to give a coordinate expansion of
$\bar{h}_{\mu \nu}^{\text{S}}$ from the covariant expansion.
The Synge world function can be written as
\begin{eqnarray}
  \sigma (x, \bar{x}) & = & \frac{1}{2} g_{\alpha \beta} (x) \Delta x^{\alpha}
  \Delta x^{\beta} + A_{\alpha \beta \gamma} (x) \Delta x^{\alpha} \Delta
  x^{\beta} \Delta x^{\gamma}\nn\\
  &  & + B_{\alpha \beta \gamma \delta} (x) \Delta x^{\alpha} \Delta
  x^{\beta} \Delta x^{\gamma} \Delta x^{\delta} + \cdots,
\end{eqnarray}
where $\Delta x^{\alpha} = x^{\alpha} - x^{\bar{\alpha}}$ and all coefficients
$A, B, \ldots$ are symmetric for all indices.
To ensure the identity
\begin{equation}
  2 \sigma  = \sigma_{\alpha} \sigma^{\alpha}
\end{equation}
holds order by order of $\Delta x^{\alpha}$,
we can obtain the coefficients.
For example, we have
\begin{equation}
  A_{\alpha \beta \gamma} (x)  =  \frac{1}{4} \partial_{(\gamma} g_{\alpha\beta)}
\end{equation}

To improve the numerical stability related to the periodicity of the azimuthal coordinate $\phi$,
we can introduce the angular variables $(Q, R)$
\begin{equation}
  \Delta \phi  =  2 \arcsin Q = \arcsin R,
\end{equation}
and write all $\Delta \phi^{2 n}$ terms into function of $Q$,
while all $\Delta \phi^{2 n+ 1}$ terms into function of $Q, R$.
\begin{eqnarray}
  \Delta \phi & = & R \left( 1 + \frac{2}{3} Q^4 \right) +\mathcal{O} (\Delta
  \phi^6),\\
  \Delta \phi^2 & = & 4 Q^2 + \frac{4}{3} Q^4 +\mathcal{O} (\Delta \phi^6),
\end{eqnarray}
As pointed out in \cite{Wardell:2011gb},
the variables $(Q, R)$ have a better behavior on numerical stability.

There exists a similar expansion for the parallel transport $g_{\phantom{\alpha}\alpha}^{\bar{\mu}}$,
\begin{equation}
  g_{\phantom{\alpha} \alpha}^{\bar{\mu}}  (x, \bar{x})  =
  \delta_{\phantom{\alpha} \alpha}^{\mu} + \Gamma_{\phantom{\alpha} \alpha
  \beta}^{\mu} (\bar{x}) \Delta x^{\beta} + B_{\phantom{\alpha} \alpha \beta
  \gamma}^{\mu} (\bar{x}) \Delta x^{\beta} \Delta x^{\gamma} + \cdots
\end{equation}
with the detailed calculation procedure for the coefficients and the expressions
for the first few orders can be found in Sec III.V of \cite{Haas:2006ne}.

After we have these coordinate expansions,
we could obtain the $\bar{h}_{\mu \nu}^{\text{S}}$ for the \acp{EMRI} on Schwarzschild.
We choose to use the common time coordinate $t = \bar{t}$ and thus
\begin{eqnarray}
  \bar{s}^2 & = & g_{\alpha \beta} \Delta x^{\alpha} \Delta x^{\beta} +
  (u_{\alpha} \Delta x^{\alpha})^2 +\mathcal{O} (\Delta x^3)\nn\\
  & = & g_{i \nocomma j} \Delta x^i \Delta x^j + (u_{\alpha} \Delta
  x^{\alpha})^2 +\mathcal{O} (\Delta x^3)
\end{eqnarray}
will always be positive except on the points where the secondary is located.

We did these calculations with xACT. The results of leading and sub-leading order can be written in the form as
\begin{flalign}\label{punc}
  \bar{h}^{\mathcal{P}}_{\alpha\beta} & = \frac{\mu}{\lambda} \frac{A_{\alpha\beta}}{\rho} + \mu \left[ \frac{B_{\alpha\beta}}{\rho} +
  \frac{C_{\alpha\beta}}{\rho^3} \right] + \mathcal{O} (\lambda)&\notag\\
  & =  \frac{\mu}{\lambda\rho}A_{\alpha\beta}
  +\frac{\mu}{\rho}B^{i}_{\alpha\beta}\Delta x^i
  +\frac{\mu}{\rho^3}\left(C^{rr\phi}_{\alpha\beta}\Delta r\Delta r\Delta\phi\right.&\notag\\
  &+C^{r\theta\theta}_{\alpha\beta}\Delta r\Delta\theta\Delta\theta
  +C^{r\phi\phi}_{\alpha\beta}\Delta r\Delta\phi\Delta\phi&\notag\\
  &\left.+C^{\theta\theta\phi}_{\alpha\beta}\Delta\theta\Delta\theta\Delta\phi
  +C^{\phi\phi\phi}_{\alpha\beta}\Delta\phi\Delta\phi\Delta\phi\right)
  + \mathcal{O} (\lambda)&
\end{flalign}
where
\begin{eqnarray}
  \rho^2&=&\frac{r[r-2M+r(u^r)^2]}{(r-2M)^2}\Delta r^2+r^2\Delta\theta^2\nonumber\\
  &  &+r^2[1+r^2(u^\phi)^2)\Delta\phi^2+\frac{2r^3u^ru^\phi}{r-2M}\Delta r\Delta\phi
\end{eqnarray}
is the leading order term of the coordinate expansion for $\bar{s}^2$,
and the $u^{\alpha}$ are the functions in \eqref{ur} $\sim$\eqref{uphi}.
Since the explicit form is very complicated, we left it in Appendix B.

To verify the reliability of our calculations, we compared the puncture field presented here with known results. It matches with the Schwarzschild quasi-circular orbit puncture field given in \cite{Wardell:2015ada}, and the leading order Schwarzschild generic orbit puncture field given in \cite{zhang2025generic}.

\subsection{Harmonic Decomposition of Field Equations}

To solve the field equations by numerical computation,
we usually decompose the equations on a tensor basis,
and express them in terms of mode components.

A common choice for Schwarzchlid background is the
Barack-Lousto-Sago (BLS) tensor harmonic basis \cite{Barack:2005nr,Barack:2007tm}.
Then $h_{\mu \nu}^{\mathcal{P}}$ can be decomposed into
\begin{eqnarray}
  \bar{h}_{i, \ell m}^{\mathcal{P}} & = & \frac{r}{a_{i, \ell} \kappa_i}
  \int^{2 \pi}_0 \mathd \phi \int^{\pi}_0 \mathd \theta \sin \theta\nn\\
  &&\times\eta^{\alpha \mu} \eta^{\beta \nu} Y_{\mu \nu}^{i, \ell m \ast} (r, \theta,
  \phi)  \bar{h}_{\mu \nu}^{\mathcal{P}}
\end{eqnarray}
Where $Y_{\mu \nu}^{i, \ell m} (r, \theta, \phi)$ are the BLS basis
and $a_{i,\ell}$ is the normalization factors.
we have $i = 1, 2, \ldots 10$ since there are ten independent components for 2nd-order symmetric tensors.
The defination of $a_{i, \ell}$, $\kappa_i$ and $\eta^{\alpha \mu}$
can be found in Appendix B of \cite{Miller:2020bft}.
Then, we can introduce a gauge-damping operator
\begin{equation}
  \breve{E}_{\mu \nu} [\bar{h}]  =  - 2 G_{\mu \nu}^{(1)} [h] - \frac{4
  M}{r^2}  [\partial_{(\mu} t]  \breve{Z}_{\nu )} [\bar{h}]
\end{equation}
where $\breve{Z}_{\mu} = (Z_r, 2Z_r, Z_{\theta}, Z_{\phi})$.
This operator will be helpful to decouple the field equation,
and the Einstein equations \eqref{G10}${\sim}$\eqref{G102} will become
\begin{eqnarray}
  \breve{E}_{\mu \nu}^{(0)} [\bar{h}^{(1)}] & = & - 16 \pi T^{(1)}_{\mu \nu}
  \label{Emn} \\
  \breve{E}_{\mu \nu}^{(0)} [\bar{h}^{(2)}] & = & - 16 \pi T^{(2)}_{\mu \nu} +
  2 \breve{G}_{\mu \nu}^{(2, 0)} [h^{(1)}, h^{(1)}]\nn\\
  && - \breve{E}_{\mu  \nu}^{(1)} [\bar{h}^{(1)}] \label{Emn0}
\end{eqnarray}
Where the $\breve{E}_{\mu \nu}^{(n)}$  are defined the same as $G_{\mu \nu}^{(n, j)}$,
and $\breve{G}_{\mu \nu}^{(n, j)}$ is $G_{\mu \nu}^{(n, j)}$ under the Lorenz condition $Z_{\alpha} = 0$.
Then we can substitute the residual field $h_{\mu \nu}^{\mathcal{R} (n)}$
and get the effective source term $S^{\mathrm{eff} (n)}_{\mu\nu}$:
\begin{eqnarray}
  S^{\mathrm{eff} (1)}_{\mu \nu}&\assign&\breve{E}_{\mu \nu}^{(0)} [\bar{h}^{\mathcal{R} (1)}] \nn\\
  & = & - \breve{E}_{\mu
  \nu}^{(0)} [\bar{h}^{\mathcal{P} (1)}] - 16 \pi T^{(1)}_{\mu \nu}
  \\
  S^{\mathrm{eff} (2)}_{\mu \nu}&\assign&\breve{E}_{\mu \nu}^{(0)} [\bar{h}^{\mathcal{R} (2)}] \nn\\
  &=& - \breve{E}_{\mu
  \nu}^{(0)} [\bar{h}^{\mathcal{P} (2)}] - 16 \pi T^{(2)}_{\mu \nu}\nn\\
  &  & + 2 \breve{G}_{\mu \nu}^{(2, 0)} [h^{(1)}, h^{(1)}] - \breve{E}_{\mu
  \nu}^{(1)} [\bar{h}^{(1)}]
\end{eqnarray}

The Lorenz condition \eqref{Z0} has a similarly expansion
\begin{eqnarray}
  &&Z_{\mu}^{(0)} [\bar{h}^{(1)}] =  0 ;\\
  &&Z_{\mu}^{(0)} [\bar{h}^{(2)}] + Z_{\mu}^{(1)} [\bar{h}^{(1)}] = 0.
\end{eqnarray}
The $\bar{h}_{\mu \nu}^{(n)}$ can be expand as
\begin{equation}
  \bar{h}_{\mu \nu}^{(n)}  =  \sum_{i, \ell m, k} \frac{1}{r}  \bar{h}_{i,
  \ell m, k}^{(n)}  (r, \mathcal{J}^I) a_{i, \ell} Y_{\mu \nu}^{i, \ell m} (r,
  \theta, \phi) e^{- \mathi (mq_{\phi} + kq_r)}
\end{equation}
We can also expand the source term $S^{(n)}_{\mu \nu}$ on the right hand of
\eqref{G10}${\sim}$\eqref{G102} as
\begin{equation}
  S^{(n)}_{\mu \nu}  = \sum_{i, \ell m, k} S^{(n)}_{i, \ell m, k}  (r,
  \mathcal{J}_I) Y_{\mu \nu}^{i, \ell m} (r, \theta, \phi) e^{- \mathi (mq_{\phi} + kq_r)}
\end{equation}

We have therefor separate the equations into a series of ODEs
\begin{equation}
  E^{(0)}_{i, j, \ell m}  \bar{h}^{(n)}_{j, \ell m, k} = \frac{2 M - r}{4 a_{i
  \ell}} S^{(n)}_{i, \ell m, k},
\end{equation}
and the $i, j = 1, \ldots 10$ are the tensor harmonic labels.
The explicit form of operator $E^{(0)}_{i, j, \ell m}$ can be found in
the appendix A of \cite{Miller:2023ers} with the replacements $\omega_{m}\rightarrow\omega_{m, k}$,
since we have included the eccentricity here.

Although we can deduce $\bar{h}_{i, \ell m}^{\mathcal{P}}$ to any order of $\Delta x$,
a direct numerical calculation in frequency domain is still hard to perform due to the eccentric motion.
Firstly, we can only numerically integrate $\bar{h}_{\ell m}^{\text{S}}$
to get its Fourier components $\bar{h}_{\ell m, k}^{\text{S}}$.
Secondly, as noted in \cite{Leather:2023dzj},
the Fourier components $T_{\ell m, k}$ of the source term $T_{\mu \nu}$
have a $|u^r (r) |^{- 1}$factor for which divergent at the turning point of the eccentric orbits,
makes it hard to handle for numerical computation.

Here we follow the extended effective sources (EES) method introduced in \cite{Leather:2023dzj}
to resolve the numerical obstacles arising from the eccentric orbits.
The effective source term $S_{\ell m}^{\mathrm{eff}}$ has non-smoothness at the point $r = r_p (t)$,
which leads to a slow convergent Fourier series.
We decompose the $S_{\ell m}^{\mathrm{eff}}$ into two smooth part
\begin{equation}
  S_{\ell m}^{\mathrm{eff}} =  S_{\ell m}^{\mathrm{eff} (+)} \Theta^{(+)} (t,
  r) + S_{\ell m}^{\mathrm{eff} (-)} \Theta^{(-)} (t, r).
\end{equation}
The $\Theta^{(\pm)} (t, r) = \Theta (\pm [r - r_p (t)])$ is the Heaviside function,
while $S_{\ell m}^{\mathrm{eff} (+)}$ and $S_{\ell m}^{\mathrm{eff}(-)}$ are
the Effective source term analytically extend from null infinity and horizon, respectively.
Then, we can define the corresponding Fourier components as
\begin{eqnarray}
 S_{\ell m, k}^{\mathrm{eff} (\pm)}  &=  &\frac{1}{T_r}  \int^{T_r}_0 S_{\ell
  m}^{\mathrm{eff} (\pm)} e^{\mathi (m \Omega_{\phi} + k \Omega_r) t} \mathd t\\
  \Theta^{(\pm)} (t, r) & = & \sum_{k \in \mathbb{Z}} b^{\pm}_k (r) e^{\mathi
  k \Omega_r t}\\
  b^{\pm}_0 (r) & = & \frac{1}{2} \pm \frac{4 t_p (r) - T_r}{2 T_r}\\
  b^{\pm}_k (r) & = & \pm \frac{1}{n \pi} \sin \left( 2 n \pi \frac{t_p
  (r)}{T_r} \right), k \neq 0
\end{eqnarray}
Thus the Fourier components of the extended effective source $S_{\ell m,k}^{\mathrm{eff}}$
is given by the convolution of $S_{\ell m, k}^{\mathrm{eff}(\pm)}$ and $b^{\pm}_k (r)$:
\begin{eqnarray}
  S_{\ell m, k}^{\mathrm{eff}} & = & \sum_{p \in \mathbb{Z}} [b^+_{p - k}
  S_{\ell m, k}^{\mathrm{eff} (+)} + b^-_{p - k} S_{\ell m, k}^{\mathrm{eff} (-)}].
\end{eqnarray}

\section{Two timescale expansion of the trajectory equations}\label{twoscaleparticle}

By eliminating the puncture field, we can obtain the residual part of the metric perturbation.
Then we can calculate the corresponding \ac{SF},
and accelerate the computation with two timescale expansion.

\subsection{The Osculating Orbits for Accelerated Motion}

As we described in the Sec. \ref{intro},
the geodesic equation in terms of the background metric can be written as
\begin{equation}
  \frac{\mathd z_{\text{p}}^{\mu}}{\mathd \tau^2} + \Gamma_{\alpha
  \beta}^{\mu}  \frac{\mathd z_{\text{p}}^{\mu}}{\mathd \tau}  \frac{\mathd
  z_{\text{p}}^{\mu}}{\mathd \tau} = f^{\mu},
\end{equation}
and $f^{\mu}$ is the \ac{SF} per unit mass.
The relations between the 2nd order \ac{SF} and the regular perturbation field are given in \cite{Pound:2017psq}
\begin{eqnarray}
  f^{\mu} & = & - \frac{1}{2} P^{\mu \alpha} \left(
  g_{\alpha}^{\phantom{\sigma} \sigma} - h_{\alpha}^{\text{R} \sigma} \right)
  \left( 2 \nabla_{\gamma} h_{\sigma \beta}^{\text{R}} - \nabla_{\sigma}
  h_{\gamma \beta}^{\text{R}} \right) u^{\beta} u^{\gamma} \nn\\
  &&+\mathcal{O}
  (\varepsilon^3) \nn\\
  & = & - \frac{1}{2} \varepsilon P^{\mu \alpha} \left( 2 \nabla_{\gamma}
  h_{\sigma \beta}^{\text{R} (1)} - \nabla_{\sigma} h_{\gamma \beta}^{\text{R}
  (1)} \right) u^{\beta} u^{\gamma} \nn\\
  &  & + \frac{1}{2} \varepsilon^2 P^{\mu \alpha} u^{\beta} u^{\gamma} \times
  \left[ h_{\alpha}^{\text{R$(1)$} \sigma} \left( 2 \nabla_{\gamma} h_{\sigma
  \beta}^{\text{R} (1)} - \nabla_{\sigma} h_{\gamma \beta}^{\text{R} (1)}
  \right) \right. \nn\\
  &  & \left. - \left( 2 \nabla_{\gamma} h_{\sigma \beta}^{\text{R} (2)} -
  \nabla_{\sigma} h_{\gamma \beta}^{\text{R} (2)} \right) \right] +\mathcal{O}
  (\varepsilon^3) \label{selfo}
\end{eqnarray}
where $P^{\mu \alpha} = g^{\mu \alpha} +u^{\mu} u^{\alpha}$.
In the numerical calculation,
we can replace $h_{mu\nu}^{\text{R}}$ with $h_{\mu\nu}^{\mathcal{R}}$.

The normalization condition $u_{\alpha} u^{\alpha} = - 1$ will led to
\begin{equation}
  \frac{\mathd}{\mathd \tau}  (u_{\alpha} u^{\alpha})  = 2u_{\alpha} f^{\alpha}= 0,
\end{equation}
and thus $f^{\alpha}$ have only 3 independent components.
Moreover, for an \ac{EMRI} system with bounded eccentric orbit on Schwarzschild spacetime,
we can set $f^{\theta} \equiv 0$ by symmetry, left only 2 independent components.
Then we will have
\begin{equation}
  - \left( 1 - \frac{2 M}{r} \right)  \frac{\mathd t}{\mathd \tau} f^t
  + \left( 1 - \frac{2 M}{r} \right)^{- 1}  \frac{\mathd r}{\mathd \tau} f^r +
  r^2  \frac{\mathd \varphi}{\mathd \tau} f^{\varphi}=0.
\end{equation}
We will choose $f^r$ and $f^{\varphi}$ as independent components, and $f^t$ can be deduced.

The real trajectory of the secondary will gradually deviate from geodesics in Schwarzschild spacetime.
A suitable method to describe such system is the
osculating orbits method \cite{Pound:2007th} in celestial mechanics.

For a given real trajectory $z^{\alpha} (\tau)$,
we can find geodesics tangent to it at each proper time $\tau$.
This geodesic is denoted by $z^{\alpha}_{\mathrm{G}} (I^A, \tau)$,
where $I_A$ are the parameters required to determine the geodesic and the secondaries' location.
Thus we can find a unique geodesic that is tangent to the real trajectory on each point of the trajectory.

The osculating geodesic method was first introduced to a generic Schwarzschild EMRI  by \cite{Pound:2007th}, while our derivation presented here draws from Section 6.1 of \cite{Pound:2021qin}.

For the bounded eccentric orbits on Schwarzschild spacetime,
we can choose $I^A = \{p, e, \chi, t, \varphi\}$,
where the first two parameters $I^a = \{p, e\}$ are the conserved quantities to determine a geodesic.
In the following we will transform the equations of the coordinates to the equations of $I^a$.

We can define $\psi^\alpha= \left(t,\chi,\varphi,\theta =\frac{\pi}{2}\right)$,
and the corresponding frequencies are denoted as
\begin{equation}
  \frac{\mathd \psi^{\alpha}}{\mathd \tau}  =  \mathfrak{f}^{\alpha}  (I^A,
  \chi) ,~~~~~~
  \frac{\mathd \psi^{\alpha}}{\mathd t}  =  \mathfrak{F}^{\alpha} (I^A,
  \chi) .
\end{equation}
Then the osculating conditions are
\begin{eqnarray}
  z^{\alpha} (\tau) & = & z^{\alpha}_G (I^a [\tau], \psi^{\alpha} [\tau]),
  \label{77}\\
  \frac{\mathd z^{\alpha} (\tau)}{\mathd \tau} & = &
  \dot{z}^{\alpha}_{\mathrm{G}} (I^a [\tau], \psi^{\alpha} [\tau]) .
  \label{88}
\end{eqnarray}
Note that $\dot{z}^{\alpha}_{\mathrm{G}}$ here is denoted as
the derivative to the proper time $\tau$ with fixed $I^a$
\begin{equation}
  \dot{z}^{\alpha}_{\mathrm{G}} (I^a [\tau], \psi^{\alpha} [\tau])  =
  \frac{\partial \dot{z}^{\alpha}_{\mathrm{G}}}{\partial \psi^{\rho}}
  \frac{\mathd \psi^{\rho}}{\mathd \tau}
\end{equation}
By differentiating Eq.\eqref{77}, we have
\begin{eqnarray}
  \frac{\mathd z^{\alpha} (\tau)}{\mathd \tau} & = & \frac{\mathd}{\mathd
  \tau} [z^{\alpha}_G (I^a [\tau], \psi^{\alpha} [\tau])]\nn\\
  & = & \frac{\partial z^{\alpha}_G}{\partial I^a}  \frac{\mathd I^a}{\mathd
  \tau} + \frac{\partial z^{\alpha}_G}{\partial \psi^{\rho}}  \frac{\mathd
  \psi^{\rho} [\tau]}{\mathd \tau}.
\end{eqnarray}
Combine with Eq.\eqref{88}, we get
\begin{equation}
  \frac{\partial z^{\alpha}_G}{\partial I^A}  \frac{\mathd I^a [\tau]}{\mathd
  \tau} + \frac{\partial z^{\alpha}_G}{\partial \psi^{\rho}}  \left(
  \frac{\mathd \psi^{\rho} [\tau]}{\mathd \tau} - \mathfrak{f}_{\rho} (I^a,
  \chi) \right) = 0
\end{equation}
According to the derivative of Eq.\eqref{88}, we will have
\begin{eqnarray}
  f^\alpha&=&\frac{\mathd^2}{\mathd \tau^2} z^{\alpha} - \frac{\partial^2}{\partial
  \tau^2} z_G^{\alpha}\nn\\
  & = & \frac{\partial
  \dot{z}^{\alpha}_{\mathrm{G}}}{\partial I^A}  \frac{\mathd I^a}{\mathd \tau}
  + \frac{\partial \dot{z}^{\alpha}_{\mathrm{G}}}{\partial \psi^{\rho}}
  \frac{\mathd \psi^{\rho}}{\mathd \tau} - \frac{\partial
  \dot{z}^{\alpha}_{\mathrm{G}}}{\partial \psi^{\rho}}  \mathfrak{f}_{\rho}
  (I^a, \chi)\nn\\
  & = & \frac{\partial \dot{z}^{\alpha}_{\mathrm{G}}}{\partial I^a}
  \frac{\mathd I^a}{\mathd \tau} + \frac{\partial
  \dot{z}^{\alpha}_{\mathrm{G}}}{\partial \psi^{\rho}}  \left[ \frac{\mathd
  \psi^{\rho}}{\mathd \tau} - \mathfrak{f}_{\rho} (I^a, \chi) \right]
\end{eqnarray}

For Schwarzschild metric, we will have the term
\begin{eqnarray}
  \delta \mathfrak{f}_{\chi} & = & \frac{\mathd \chi}{\mathd \tau} -
  \mathfrak{f}_{\chi}  (I^a, \chi)\nn\\
  & = & \left( \frac{\partial r}{\partial \chi} \right)^{- 1}  \left(
  \frac{\partial r}{\partial p}  \frac{\mathd p}{\mathd \tau} + \frac{\partial
  r}{\partial e}  \frac{\mathd e}{\mathd \tau} \right)\nn\\
  & = & - \frac{1 + e \cos \chi}{ep \sin \chi} \left( \frac{\mathd p}{\mathd
  \tau} \right) + \frac{\cos \chi}{e \sin \chi} \left( \frac{\mathd e}{\mathd
  \tau} \right)
\end{eqnarray}
and
\begin{eqnarray}
  \frac{\mathd p}{\mathd \tau} & = & \frac{\mathcal{L} (e, \varphi) f^r
  -\mathcal{L} (e, r) f^{\varphi}}{\mathcal{L} (p, r) \mathcal{L} (e, \varphi)
  -\mathcal{L} (e, r) \mathcal{L} (p, \varphi)}\\
  \frac{\mathd e}{\mathd \tau} & = & \frac{-\mathcal{L} (p, \varphi) f^r
  +\mathcal{L} (p, r) f^{\varphi}}{\mathcal{L} (p, r) \mathcal{L} (e, \varphi)
  -\mathcal{L} (e, r) \mathcal{L} (p, \varphi)}
\end{eqnarray}
with
\begin{equation}
    \mathcal{L} (I^a, f) = \frac{\partial \dot{f}}{\partial I^a} - \left(
   \frac{\partial r}{\partial \chi} \right)^{- 1} \frac{\partial r}{\partial
   I^a}  \frac{\partial \dot{f}}{\partial \chi}
\end{equation}
Multiply by $\mathd \tau/\mathd t$, we have
\begin{eqnarray}
  \frac{\mathd p}{\mathd t} & = & \mathfrak{c}_r^{(p)} f^r
  +\mathfrak{c}_{\varphi}^{(p)} f^{\varphi}  \label{alf}\\
  \frac{\mathd e}{\mathd t} & = & \mathfrak{c}_r^{(e)} f^r
  +\mathfrak{c}_{\varphi}^{(e)} f^{\varphi}  \label{bt}\\
  \frac{\mathd \chi}{\mathd t} & = & \mathfrak{F}^r (I^a, \chi) \nonumber\\&&- \frac{1 + e
  \cos \chi}{ep \sin \chi} \left( \frac{\mathd p}{\mathd t} \right) +
  \frac{\cos \chi}{e \sin \chi} \left( \frac{\mathd e}{\mathd t} \right) \\
  \frac{\mathd \phi}{\mathd t} & = & \mathfrak{F}^{\phi} (I^a, \chi)
  \label{omg}
\end{eqnarray}
Eq. \eqref{alf}${\sim}$ Eq. \eqref{omg} are equations for accelerated motion,
and the $\mathfrak{c}_{\alpha}^{(a)}$ are defined as
\begin{eqnarray*}
  \mathfrak{c}_r^{(p)} & = & - 2 e p \sin \chi \sqrt{\frac{p - 6 - 2 e \cos
  \chi}{(p - 2)^2 - 4 e^2}} \times\\
  &  & \frac{(p - e^2 - 3)  (p - 2 e \cos \chi - 2)}{(p - 2 e - 6)  (p + 2 e
  - 6) };\\
  &  & \\
  \mathfrak{c}_{\varphi}^{(p)} & = & \sqrt{\frac{p}{(p - 2)^2 - 4 e^2}} \frac{
  (p - e^2 - 3) (p - 2 e \cos \chi - 2) }{(p - 2 e - 6)  (p + 2 e - 6)}
  \times\\
  &  & \frac{p M (p - 6 - 2 e \cos \chi)  (2 p - e^2 [\cos 2 \chi + 1] - 6)}{
  (e \cos \chi + 1)^2};\\
  &  & \\
  \mathfrak{c}_r^{(e)} & = & \sin \chi \sqrt{\frac{p - 6 - 2 e \cos \chi}{(p -
  2)^2 - 4 e^2}} \times\\
  &  & \frac{(p - e^2 - 3)  (p - 2 e^2 - 6)  (p - 2 - 2 e \cos \chi)}{(p - 2
  e - 6)  (p + 2 e - 6) };\\
  &  & \\
  \mathfrak{c}_{\varphi}^{(e)} & = &  \sqrt{\frac{p}{(p - 2)^2 - 4 e^2}}
  \frac{M (p - e^2 - 3)}{2 (p - 2 e - 6)  (p + 2 e - 6) } \times\\
  &  & \frac{ (p - 2 e \cos \chi - 2)}{ (e \cos \chi + 1)^2}
  [\nobracket (p - 2 e^2 - 6) \times\\
  &  & [(4 p - 3 e^2 - 12) \cos \chi + e^2 \cos 3 \chi - e (p - 6) \cos 2
  \chi]\\
  &  & + e (p (3 p - 32) - 2 e^2 (p - 10) + 60) \nobracket].
\end{eqnarray*}
these equations can be solved numerically,
once we obtained the $f^{\alpha}$ derived from the field equations.

\subsection{The two timescale expansion for the trajectory equations}

In order to reduce the computation cost,
we also need to do the two timescale expansion on the equation of the trajectory.
According to the Fourier expansion
\begin{equation}
  F (\chi)  = \sum_{k \in \mathbb{Z}} F_k \exp [- \mathi k q_r] \label{fly}
\end{equation}
We can get
\begin{eqnarray}
  \chi & = & q_r + A_r + \Delta \psi_r (q_r),\\
  \phi & = & q_{\phi} + A_{\phi} + \Delta \psi_{\phi} (q_r),
\end{eqnarray}
where
\begin{eqnarray}
  \Delta \psi_{\alpha} (q_r) & = & \sum_{k \neq 0} \frac{\mathfrak{F}_{\alpha,
  k}}{- \mathi k \Omega_r} \exp [- \mathi kq_r]\\
  A_{\alpha} & = & - \Delta \psi_{\alpha} (0)
\end{eqnarray}

To clarify the effects of self-force of different orders, We can expand $f^{\alpha}$ as a serie of $\varepsilon$,
\begin{eqnarray*}
  f^{\alpha} & = & \varepsilon f^{\alpha (1)} + \varepsilon^2 f^{\alpha (2)} +
  \cdots
\end{eqnarray*}
the right-hand sides of
$\mathrm{Eq} . \eqref{alf} \sim \mathrm{Eq} . \eqref{omg}$ can be written in the form
\begin{eqnarray}
  \frac{\mathd q_r}{\mathd t} & = & \Omega_r [I_c (\tilde{t})] + \varepsilon
  g_{\alpha}  (q_r, I_c, \tilde{t}, \varepsilon) \nn\\
  & = & \Omega_r [I_c (\tilde{t})] + \sum_{n = 1}^{+ \infty} \varepsilon^n
  g_r^{(n)}   \label{qq}\\
  \frac{\mathd I_a}{\mathd t} & = & \varepsilon G_a (q_r, I_c, \tilde{t},
  \varepsilon) = \sum_{n = 1}^{+ \infty} \varepsilon^n G_a^{(n)}  \label{II}\\
  \frac{\mathd q_{\phi}}{\mathd t} & = & \Omega_{\phi} [I_c (\tilde{t})]
  \label{qf}
\end{eqnarray}
where
\begin{eqnarray*}
  G_a^{(n)} & = &  \mathfrak{c}_{\alpha}^{(a)} f^{\alpha(n)},
\end{eqnarray*}
and
\begin{eqnarray*}
  g_r^{(n)} & = &  \frac{\Omega_r f^{\alpha(n)}}{ \mathfrak{F}^{\alpha}(I^a, \chi)}\left[- \frac{1 + e
  \cos \chi}{ep \sin \chi} G_p^{(n)} +
  \frac{\cos \chi}{e \sin \chi} G_e^{(n)}\right].
\end{eqnarray*}

Note that since $q_{\phi}$ does not appear on the right-hand side,
we can leave Eq.\eqref{qf} alone and focus on
solving the two coupled equations Eq.\eqref{qq} and Eq.\eqref{II}.All effects of $n$-order self-force $f^{\alpha(n)}$ are therefore fully represented by $G_a^{(n)}$ and $g_r^{(n)}$.
The detailed expressions of  $G_a^{(n)},g_r^{(n)}$ needed for 1PA calculation are given in Appendix \ref{app:tte}.

Following the definitions in \cite{Hinderer:2008dm},
we define the short timescale variable $\Psi =q_r +\mathcal{O} (\varepsilon)$ as
\begin{eqnarray}
  \frac{\mathd \Psi}{\mathd t} & = & \Omega_r [I_c (\tilde{t})]\nn\\
  & = & \Omega^{(0)}  (\varepsilon t) + \varepsilon \Omega^{(1)}
  (\varepsilon t) +\mathcal{O} (\varepsilon^2)
\end{eqnarray}
Integrating the above equation with respect to $\tilde{t} = \varepsilon t$,
we can get the mass ratio expansion of $q_{\alpha}$,
\begin{equation}
  \Psi (\tilde{t}, \varepsilon)  =  \frac{1}{\varepsilon} \Psi^{(0)}
  (\tilde{t}) + \Psi^{(1)} (\tilde{t}) +\mathcal{O} (\varepsilon),
\end{equation}
where $\frac{1}{\varepsilon} \Psi^{(0)} (\tilde{t})$ term is the 0PA term,
and $\Psi^{(1)} (\tilde{t})$ is the 1PA term.

The action-angle variables and geodesic parameters
as functions of short timescale $\Psi$ and long timescale $\tilde{t}$
can be expanded as
\begin{eqnarray}
  q_r & = & \sum_{n = 0}^{+ \infty} \varepsilon^n q_r^{(n)} (\Psi,
  \tilde{t})\\
  I_a & = & \sum_{n = 0}^{+ \infty} \varepsilon^n I_a^{(n)} (\Psi, \tilde{t})
\end{eqnarray}
Since $\Psi$ describes the fast periodic motion, we have
\begin{eqnarray}
  q_r^{(0)} (\Psi + 2 \pi, \tilde{t}) & = & q_r^{(0)} (\Psi, \tilde{t}) + 2
  \pi.\\
  q_r^{(n)} (\Psi + 2 \pi, \tilde{t}) & = & q_r^{(n)} (\Psi, \tilde{t}) ,~~~~~~
 n > 0.\\
  I_a^{(n)} (\Psi + 2 \pi, \tilde{t}) & = & I_a^{(n)} (\Psi, \tilde{t}) .
\end{eqnarray}
Then $q_r$ can be written as
\begin{eqnarray}
  q_r (\Psi, \tilde{t}) & = & \Psi + \Delta q_r (\Psi, \tilde{t}) + \varepsilon q_r^{(1)}
  (\Psi, \tilde{t}) + \cdots \nn\\
  & = & \frac{1}{\varepsilon} \Psi^{(0)} (\tilde{t}) + [\Psi^{(1)}
  (\tilde{t}) + \Delta q_r (\Psi, \tilde{t})] \nn\\
  &  & + \varepsilon [q_r^{(1)} (\Psi, \tilde{t}) + \Psi^{(2)} (\tilde{t})]
  +\mathcal{O} (\varepsilon^2)
\end{eqnarray}
where $\Delta q_r (\Psi, \tilde{t}) = q_r^{(0)} (\Psi, \tilde{t}) - \Psi$
is a $2 \pi$-periodic function.

The derivative of $t$ can be expanded as
\begin{eqnarray}
  \frac{\mathd}{\mathd t} F (\Psi, \tilde{t}) & = & [\Omega_r \partial_{\Psi}
  + \varepsilon \partial_{\tilde{t}}] F (\Psi, \tilde{t})\nn\\
  & = & [\Omega^{(0)} (\tilde{t}) \partial_{\Psi} + \varepsilon (\Omega^{(1)}
  (\tilde{t}) \partial_{\Psi} + \partial_{\tilde{t}})]\times\nn\\
  & &  F (\Psi, \tilde{t})
\end{eqnarray}
The left hand sides of Eq.\eqref{qq} and Eq.\eqref{II} can therefore be expanded as
\begin{eqnarray}
  \frac{\mathd q_r}{\mathd t} & = & \Omega^{(0)} (\tilde{t}) \partial_{\Psi}
  q_r^{(0)} +\\
  &  & \varepsilon [\partial_{\tilde{t}} q_r^{(0)} + \Omega^{(1)} (\tilde{t})
  \partial_{\Psi} q_r^{(0)} + \Omega^{(0)} (\tilde{t}) \partial_{\Psi}\
  q_r^{(1)}] +\mathcal{O} (\varepsilon^2)\nn\\
  \frac{\mathd I_a}{\mathd t} & = & \Omega^{(0)} (\tilde{t}) \partial_{\Psi}
  I^{(0)}_a + \varepsilon [\partial_{\tilde{t}} I_a^{(0)} + \Omega^{(1)}
  (\tilde{t}) \partial_{\Psi} I_a^{(0)}\nn\\
  && + \Omega^{(0)} (\tilde{t})
  \partial_{\Psi} I_a^{(1)}]  +\varepsilon^2 [\Omega^{(2)} (\tilde{t}) \partial_{\Psi} I^{(0)}_a +
  \widetilde{\partial_t} I_a^{(1)} +\nn\\
  && \Omega^{(1)} (\tilde{t}) \partial_{\Psi}
  I_a^{(1)} + \Omega^{(0)} (\tilde{t}) \partial_{\Psi} I_a^{(2)}]+\mathcal{O} (\varepsilon^3)
\end{eqnarray}
While the right-hand sides can be expanded as
\begin{eqnarray}
  &  & \Omega_r (\tilde{t}, \varepsilon) + \varepsilon g_r (q_r, I_c,
  \tilde{t}, \varepsilon)\nn\\
  & = & \Omega_r (I^{(0)}_c) + \varepsilon [I_b^{(1)} \partial_{I_b} \Omega_r
  (I^{(0)}_c) \nobracket \nobracket + g_r^{(1)} (q_r^{(0)}, I^{(0)}_c,
  \tilde{t})]\nn\\
  &  & +\mathcal{O} (\varepsilon^2) ;\\
  &  & \varepsilon G_a (q_r, I_c, \tilde{t}, \varepsilon)\nn\\
  & = & \varepsilon G_a^{(1)}  (q_r^{(0)}, I^{(0)}_c, \tilde{t}) +
  \varepsilon^2 [q_r^{(1)} \partial_{q_r} G_a^{(1)} (q_r^{(0)}, I^{(0)}_a,
  \tilde{t}) \nobracket\nn\\
  &  & \nobracket + I^{(1)}_b \partial_{I _b} G_a^{(1)} (q_r^{(0)},
  I^{(0)}_c, \tilde{t}) + G_a^{(2)}  (q_r^{(0)}, I^{(0)}_c, \tilde{t})]\nn\\
  &  & +\mathcal{O} (\varepsilon^3),
\end{eqnarray}

Compare the two sides order by order of $\varepsilon$, we get the two timescale equations
\begin{eqnarray}
  \Omega^{(0)} (\tilde{t}) \partial_{\Psi} I^{(0)}_a & = & 0 \label{0I} \\
  \Omega^{(0)} (\tilde{t}) \partial_{\Psi} q_r^{(0)} & = & \Omega_r
  (I^{(0)}_c) \label{0q} \\
  \Omega^{(0)} (\tilde{t}) \partial_{\Psi} I_a^{(1)} & = & G_a^{(1)}
  (q_r^{(0)}, I^{(0)}_c, \tilde{t})\nn\\
  && - \Omega^{(1)} (\tilde{t}) \partial_{\Psi}
  I_a^{(0)} - \partial_{\tilde{t}} I_a^{(0)} \label{1I} \\
  \Omega^{(1)} (\tilde{t}) \partial_{\Psi} q_r^{(0)} & = & I^{(1)}_b
  \partial_{I _b} \Omega_r (I^{(0)}_c) + g_r^{(1)}  (q_r^{(0)}, I^{(0)}_c,
  \tilde{t}) \nn\\
  &  & - \Omega^{(0)} (\tilde{t}) \partial_{\Psi} q_r^{(1)} -
  \partial_{\tilde{t}} q_r^{(0)} \label{1q} \\
  \Omega^{(2)} (\tilde{t}) \partial_{\Psi} I^{(0)}_a & = & q_r^{(1)}
  \partial_{q_r} G_a^{(1)} (q_r^{(0)}, I^{(0)}_c, \tilde{t}) \nn\\
  &  & + I^{(1)}_b \partial_{I _b} G_a^{(1)} (q_r^{(0)}, I^{(0)}_c,
  \tilde{t}) \nn\\
  &  & + G_a^{(2)}  (q_r^{(0)}, I^{(0)}_c, \tilde{t}) - \Omega^{(0)}
  (\tilde{t}) \partial_{\Psi} I_a^{(2)} \nn\\
  &  & - \Omega^{(1)} (\tilde{t}) \partial_{\Psi} I_a^{(1)} -
  \partial_{\tilde{t}} I_a^{(1)} \label{2I} \\
  & \vdots &  \nn
\end{eqnarray}
Higher order equations are not needed for 1PA evolutions.
We used the Einstein summation convention that
$I^{(1)}_b \partial_{I_b} = p^{(1)}(\tilde{t}) \partial_p +e^{(1)}(\tilde{t}) \partial_e$.

\subsection{The twotimescale expansion in frequency domain}

To simplify the expression and calculate in frequency domain,
we introduce some new notation here.
For any 2$\pi$-periodic function $F (\Psi)$, by perform Fourier decompose
\begin{equation}
  F (\Psi)  =  \sum_{k \in \mathbb{Z}} F_k e^{\mathi k \Psi},
\end{equation}
with
\begin{equation}
  F_k  =  \frac{1}{2 \pi}  \int^{2 \pi}_0 F (\Psi) e^{- \mathi k \Psi}
  \mathd \Psi,
\end{equation}
we can decompose $F (\Psi)$ into an oscillatory component $\hat{F} (\Psi)$
and the averaged value$\langle F \rangle$ as
\begin{eqnarray}
  F (\Psi) & = & \hat{F} (\Psi) + \langle F \rangle ,\\
  \langle F \rangle = F_0 & = & \frac{1}{2 \pi}  \int^{2 \pi}_0 F (\Psi)
  \mathd \Psi,\\
  \hat{F} (\Psi) & = & \sum_{k \neq 0} F_k e^{\mathi k \Psi} .
\end{eqnarray}
Here we define the ``anti-derivative'' operator $\mathcal{I}$ for oscillatory component as
\begin{equation}
  \mathcal{I} \hat{F} (\Psi)  =  \sum_{k \neq 0} \frac{F_k}{\mathi k}
  e^{\mathi k \Psi},
\end{equation}
Then, we will have the following identities:
\begin{eqnarray}
  \langle F (\Psi) G (\Psi) \rangle & = & \langle \hat{F} (\Psi) \hat{G}
  (\Psi) \rangle + \langle F \rangle \langle G \rangle\\
  \frac{\mathd \mathcal{I} \hat{F}}{\mathd \Psi} =\mathcal{I} \frac{\mathd
  \hat{F}}{\mathd \Psi} & = & \hat{F} (\Psi)\\
  \langle [\mathcal{I} \hat{F} (\Psi)] \hat{G} (\Psi) \rangle & = & - \langle
  [\mathcal{I} \hat{G} (\Psi)] \hat{F} (\Psi) \rangle
\end{eqnarray}
Eq.\eqref{0I} shows that $I^{(0)}_a (\Psi, \tilde{t}) = I^{(0)}_a (\tilde{t})$,
which means that the geodesic parameters will not be effected
by the short timescale oscillation at zeroth-order approximation.

Eq.\eqref{0q} gives that
\begin{equation}
  q_r^{(0)} (\Psi, \tilde{t})  =  \frac{\Omega_r [I^{(0)}_c
  (\tilde{t})]}{\Omega^{(0)} (\tilde{t})} \Psi + q_r^{(0)} (0, \tilde{t})
\end{equation}
Since $\Delta q_r (\Psi, \tilde{t}) = q_r^{(0)} (\Psi, \tilde{t}) - \Psi$
is a 2$\pi$-periodic function, we have
\begin{equation}
  \Omega^{(0)} (\tilde{t})  =  \Omega_r [I^{(0)}_c (\tilde{t})] .
\end{equation}
We can see that the zeroth-order equations is just the geodesic motions.
$q_r^{(0)} (0, \tilde{t})$ can be set to  $ 0$, and we will have
\begin{equation}
  q_r^{(0)} (\Psi, \tilde{t})  =  \Psi .
\end{equation}
The functions $F (q_r^{(0)}, I^{(0)}_a, \tilde{t})$ in Eq.\eqref{1I}$\sim$Eq.\eqref{2I}
are therefore $F (\Psi, I^{(0)}_a (\tilde{t}),\tilde{t})$,
we will omit the arguments $(\Psi, I^{(0)}_a (\tilde{t}), \tilde{t})$
in the following part for simplicity.

Now, Eq.\eqref{1I} reads as
\begin{equation}
  \Omega^{(0)} (\tilde{t}) \partial_{\Psi} I_a^{(1)}  =  G_a^{(1)} -
  \partial_{\tilde{t}} I_a^{(0)}
\end{equation}
Take average on both sides, we will have
\begin{equation}
  \partial_{\tilde{t}} I^{(0)}_a (\tilde{t})  = \langle G_a^{(1)} \rangle
  (I^{(0)}_c (\tilde{t}), \tilde{t})\label{I0a}
\end{equation}
which is the adiabatic evolution of $I^{(0)}_a (\tilde{t})$.

The remaining oscillatory component of Eq.\eqref{1I} will be
\begin{equation}
  \hat{I}_a^{(1)}  =  \frac{\mathcal{I} \hat{G}_a^{(1)}}{\Omega^{(0)}
  (\tilde{t})}
\end{equation}
We denote $\langle I_a^{(1)} \rangle$ as $\mathcal{C}_a^{(1)}
(\tilde{t})$, and the full expression is
\begin{equation}
  I_a^{(1)} =  \frac{\mathcal{I} \hat{G}_a^{(1)}}{\Omega^{(0)} (\tilde{t})}
  +\mathcal{C}_a^{(1)} (\tilde{t}) .  \label{I1a}
\end{equation}
Substitute \eqref{I1a} into \eqref{1q} and take the average, we will have
\begin{equation}
  \Omega^{(1)} (\tilde{t})  =  \mathcal{C}_a^{(1)} (\tilde{t})
  \partial_{I_a} \Omega_r (I^{(0)}_c) + \langle g^{(1)}_r\rangle
\end{equation}
and the oscillatory component
\begin{eqnarray}
  \Omega^{(0)} (\tilde{t}) \partial_{\Psi}  \hat{q}^{(1)} & = &
  \hat{g}_r^{(1)} + \hat{I}_b^{(1)} \partial_{I_b} \Omega_r (I^{(0)}_c)\\
  & = & \hat{g}_r^{(1)} + \frac{\mathcal{I} \hat{G}_b^{(1)}}{\Omega^{(0)}
  (\tilde{t})} \partial_{I_b} \Omega_r (I^{(0)}_c),
\end{eqnarray}
will give the solution of $\hat{q}_r^{(1)}$ and $q_r^{(1)}$.
\begin{eqnarray}
  \hat{q}_r^{(1)} & = & \frac{\mathcal{I} \hat{g}_r^{(1)}}{\Omega^{(0)}
  (\tilde{t})} + \frac{\mathcal{I}^2  \hat{G}_b^{(1)} \cdot \partial_{I_b}
  \Omega_r (I^{(0)}_c)}{[\Omega^{(0)} (\tilde{t})]^2} \nonumber\\
  &  &  \nonumber\\
  q_r^{(1)} & = & \hat{q}_r^{(1)} +\mathcal{Q} (\tilde{t}) = \hat{q}_r^{(1)} -
  \hat{q}_r^{(1)} (0, \tilde{t})  \label{q1r}
\end{eqnarray}

The last component to determine $\Omega^{(1)} (\tilde{t})$ is the evolution of
$\mathcal{C}^{(1)} (\tilde{t})$, which comes from the average part of Eq.\eqref{2I},
\begin{eqnarray}
  \partial_{\tilde{t}} \mathcal{C}_a^{(1)} (\tilde{t})  =  \langle q_r^{(1)}
  \partial_{\Psi} G_a^{(1)} \rangle + \langle I^{(1)}_b \partial_{I_b}
  G_a^{(1)} \rangle + \langle G_a^{(2)} \rangle  \label{tC1}
\end{eqnarray}
Substitute Eq.\eqref{I1a} and Eq.\eqref{q1r} into Eq.\eqref{tC1},
and utilizing the identities of operator $\mathcal{I}$,
we can finally obtain the equation of $\mathcal{C}_a^{(1)}$
\begin{flalign}
  \widetilde{\partial_t} \mathcal{C}_a^{(1)} (\tilde{t})  = &
  \mathcal{C}_b^{(1)} (\tilde{t}) \partial_{I_b} \langle G_a^{(1)} \rangle + \langle G_a^{(2)} \rangle &\notag\\
  &+\frac{\langle \mathcal{I} \hat{G}_b^{(1)} \cdot \partial_{I_b}
  \hat{G}_a^{(1)} \rangle - \langle \hat{g}_r^{(1)}  \hat{G}_a^{(1)}
  \rangle}{\Omega^{(0)} (\tilde{t})}& \label{C1a}.\
\end{flalign}
Note that the $ \langle G_a^{(2)} \rangle$ term is the only term in 2nd order,
which stands for the dissipative part of 2nd SF $f_{\mathrm{diss}}^{(2) \mu}$.

For the evolution of $q_{\phi}$,
since $\Omega_{\phi} [I_c(\tilde{t})] = \Omega_{\phi}  [I_c (\varepsilon t)]$ is free from $q_r$,
we can set $q_{\phi} = q_{\phi} (\tilde{t})$ and expand  it as
\begin{equation}
  q_{\phi} (\tilde{t})  =  \frac{1}{\varepsilon} \Phi^{(0)} (\tilde{t}) +
  \Phi^{(1)} (\tilde{t}) +\mathcal{O} (\varepsilon^2)
\end{equation}
Now, take the average terms of Eq.\eqref{qf}, we have
\begin{eqnarray}
  \partial_{\tilde{t}} \Phi^{(0)} (\tilde{t}) & = & \Omega_{\phi} [I^{(0)}_c
  (\tilde{t})]\label{f0}\\
  \partial_{\tilde{t}} \Phi^{(1)} (\tilde{t}) & = &\mathcal{C}_b^{(1)}  \partial_{I_b}
  \Omega_{\phi} (I^{(0)}_c [\tilde{t}])\label{f1}
\end{eqnarray}

Finally, we have a series of ordinary differential equations which are independent of $\varepsilon$.
Then these equations(\eqref{I0a},\eqref{I1a},\eqref{q1r},\eqref{f0},\eqref{f1})
can be efficiently solved numerically once the magnitude of the self-force is specified.

\subsection{Numerical result}

Using the equations obtained above,
we can do numerical calculations on the evolution of the orbit.
Here we just provide a simple verification
of the validity and efficiency of our analytical formulas.
So we used the analytical formulas of the self-force
in \ac{PN} limit given by \cite{Pound:2007th}.

\begin{figure*}[htbp]
    \centering

    \begin{subfigure}[b]{0.32\textwidth}
        \includegraphics[width=\textwidth]{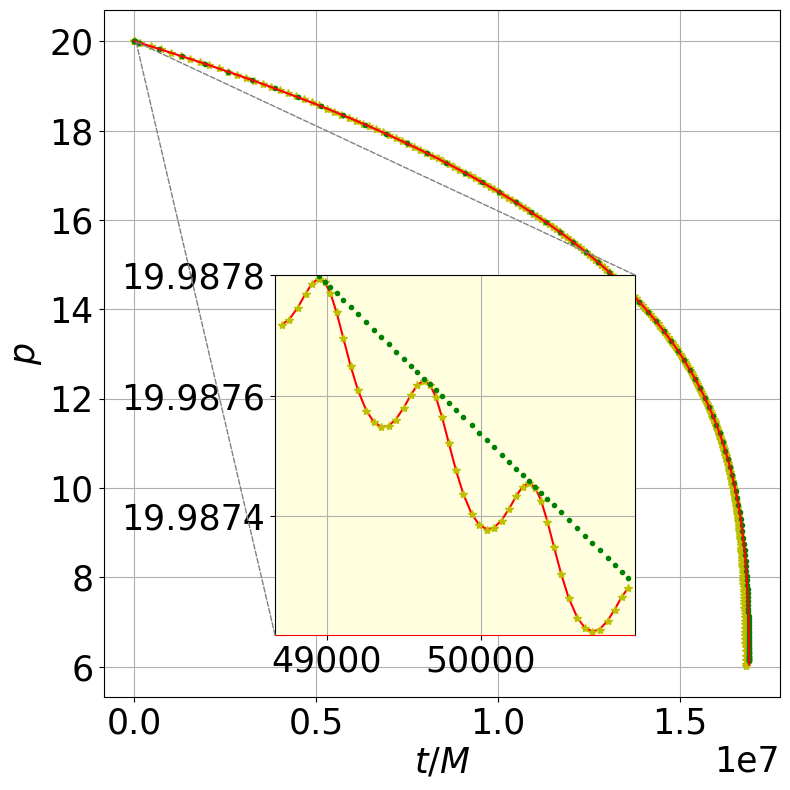}
    \end{subfigure}
    \hfill
    \begin{subfigure}[b]{0.32\textwidth}
        \includegraphics[width=\textwidth]{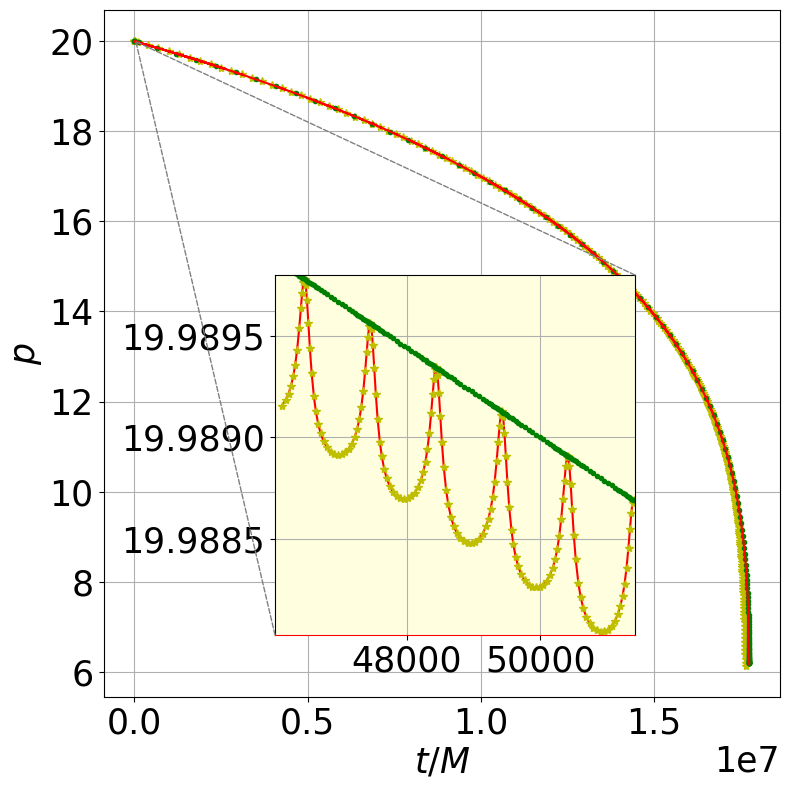}
    \end{subfigure}
    \hfill
    \begin{subfigure}[b]{0.32\textwidth}
        \includegraphics[width=\textwidth]{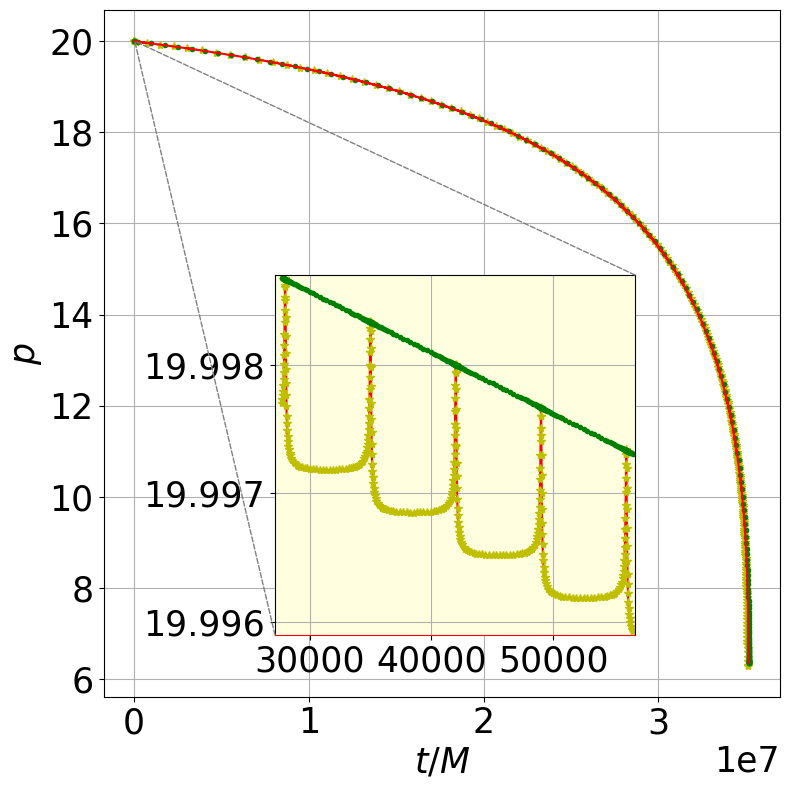}
    \end{subfigure}

    \vspace{1em}
    \begin{subfigure}[b]{0.32\textwidth}
        \includegraphics[width=\textwidth]{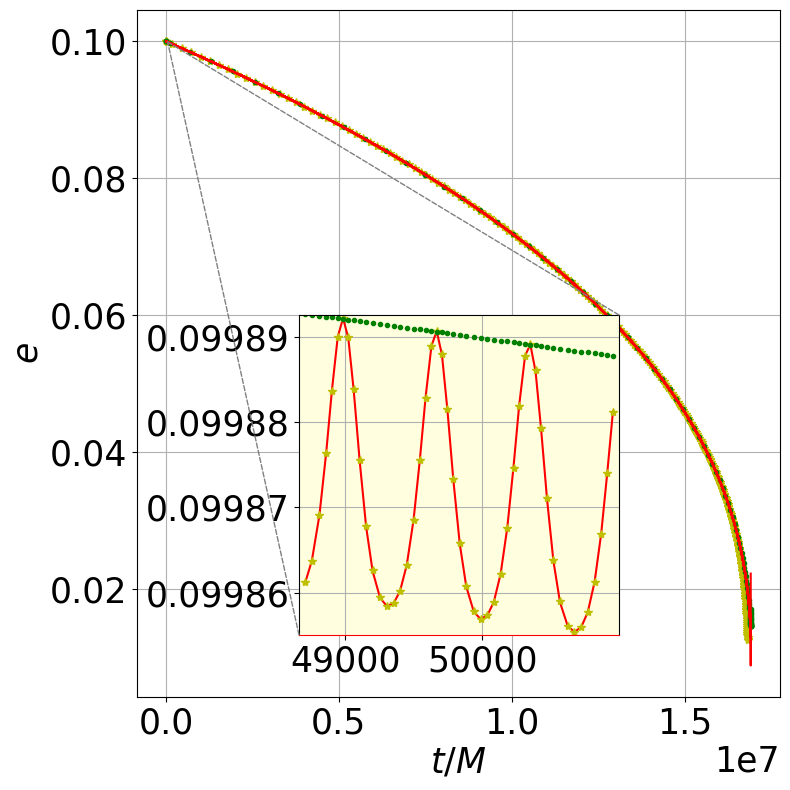}
    \end{subfigure}
    \hfill
    \begin{subfigure}[b]{0.32\textwidth}
        \includegraphics[width=\textwidth]{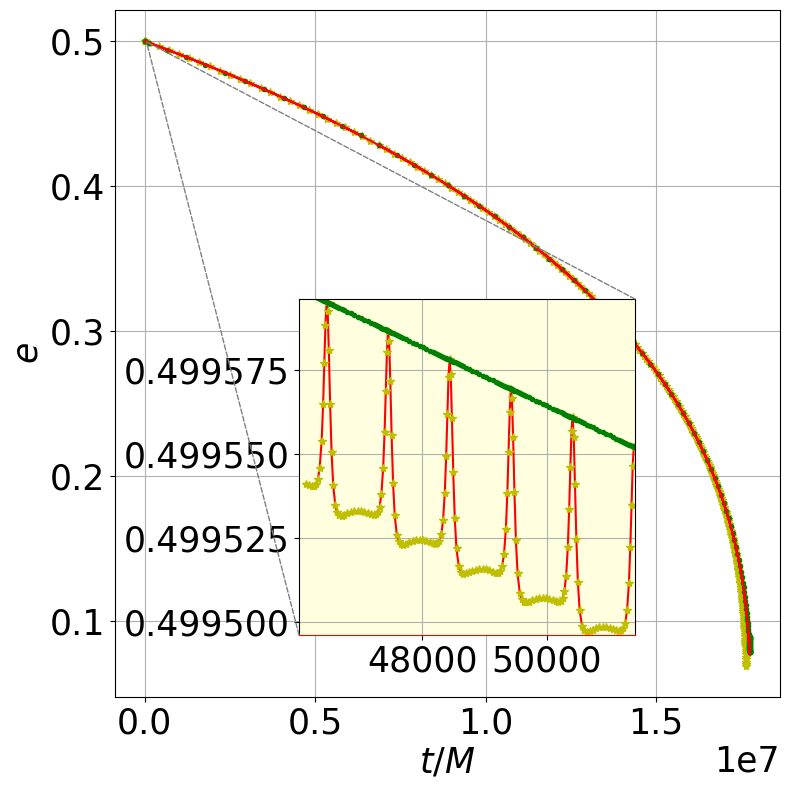}
    \end{subfigure}
    \hfill
    \begin{subfigure}[b]{0.32\textwidth}
        \includegraphics[width=\textwidth]{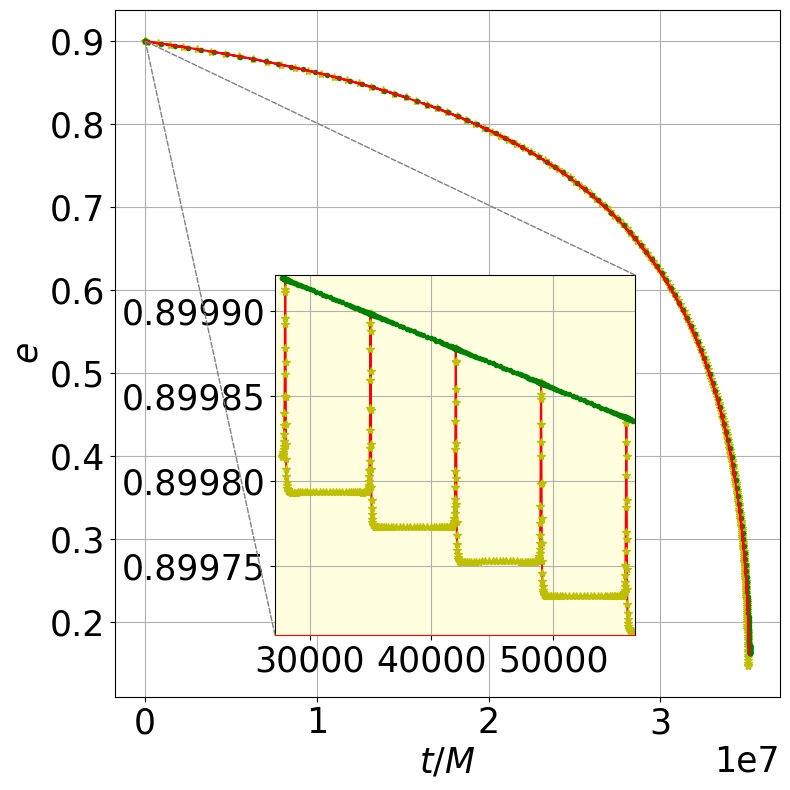}
    \end{subfigure}

    \begin{subfigure}[b]{\textwidth}
      \noindent
    Osculating geodesic:
    \begin{tikzpicture}[baseline=-0.5ex, every node/.style={inner sep=0pt}]
        \draw[red, thick] (0,0) -- (1cm,0);
    \end{tikzpicture}\quad
    Two timescale: 1PA :
    \begin{tikzpicture}[baseline=-0.5ex]
        \node[anchor=base, color={rgb,255:red,204; green,204; blue,0}, scale=1.2] at (0,-0.2) {***};
    \end{tikzpicture}\quad
    Two timescale: 0PA :
    \begin{tikzpicture}[baseline=-0.5ex]
        \node[anchor=base, color={rgb,255:red,0; green,100; blue,0}, scale=1.5] at (0,-0.2) {\scalebox{2}{$\cdots$}};
    \end{tikzpicture}
    \end{subfigure}

    \caption{\raggedright The evolution of $p$ and $e$ as functions of dimensionless time for \acp{EMRI} with $\varepsilon=10^{-4}$.
    For each figure, the initial value of $p$ is $20$, the evolution ended at the separatrix of $p = 6+2e$, the result for osculating geodesic is plotted in solid red line, the two timescale for 0PA is plotted in green dot, and the two timescale for 1PA is plotted in yellow star.
    Each columns corresponds to different initial eccentricity $e$: $e=0.1$ for the left column, $e=0.5$ for the middle column, and $e=0.9$ for the right column. We also show the zoom in comparison around $t=5\varepsilon M$.} \label{fig:orbit_p}
\end{figure*}

We calculate the evolution of $p$ and $e$,
for \acp{EMRI} with different $\varepsilon$ and initial $e$.
The initial values of $p$ are chosen to be $20$ for all the cases.
For a comparison, we do the same calculation by integrating the osculating equations directly.
On the other hand, we calculate both the 0PA and 1PA result with two timescale method.
All the calculations are implemented using adaptive step-size Runge-Kutta algorithms for numerical integration and Fast Fourier Transforms (FFTs) for Fourier coefficients.
For simplicity and rapid prototyping, all computations are performed in Python without any precomputation such as offline GSF database;
every operation is carried out online in real time.
The results for $\varepsilon=10^{-4}$ with different initial eccentricities ($e=0.1$, $0.5$, $0.9$) are shown in Fig.~\ref{fig:orbit_p}.We can find that the 0PA result can reproduce the overall evolution trend without the small-scale oscillations.
However, if we consider the 1PA order terms,
the result will have an excellent agreement with the osculating geodesic integration.

\begin{figure*}[htbp]
    \centering
    \begin{subfigure}[b]{0.32\textwidth}
        \includegraphics[width=\textwidth]{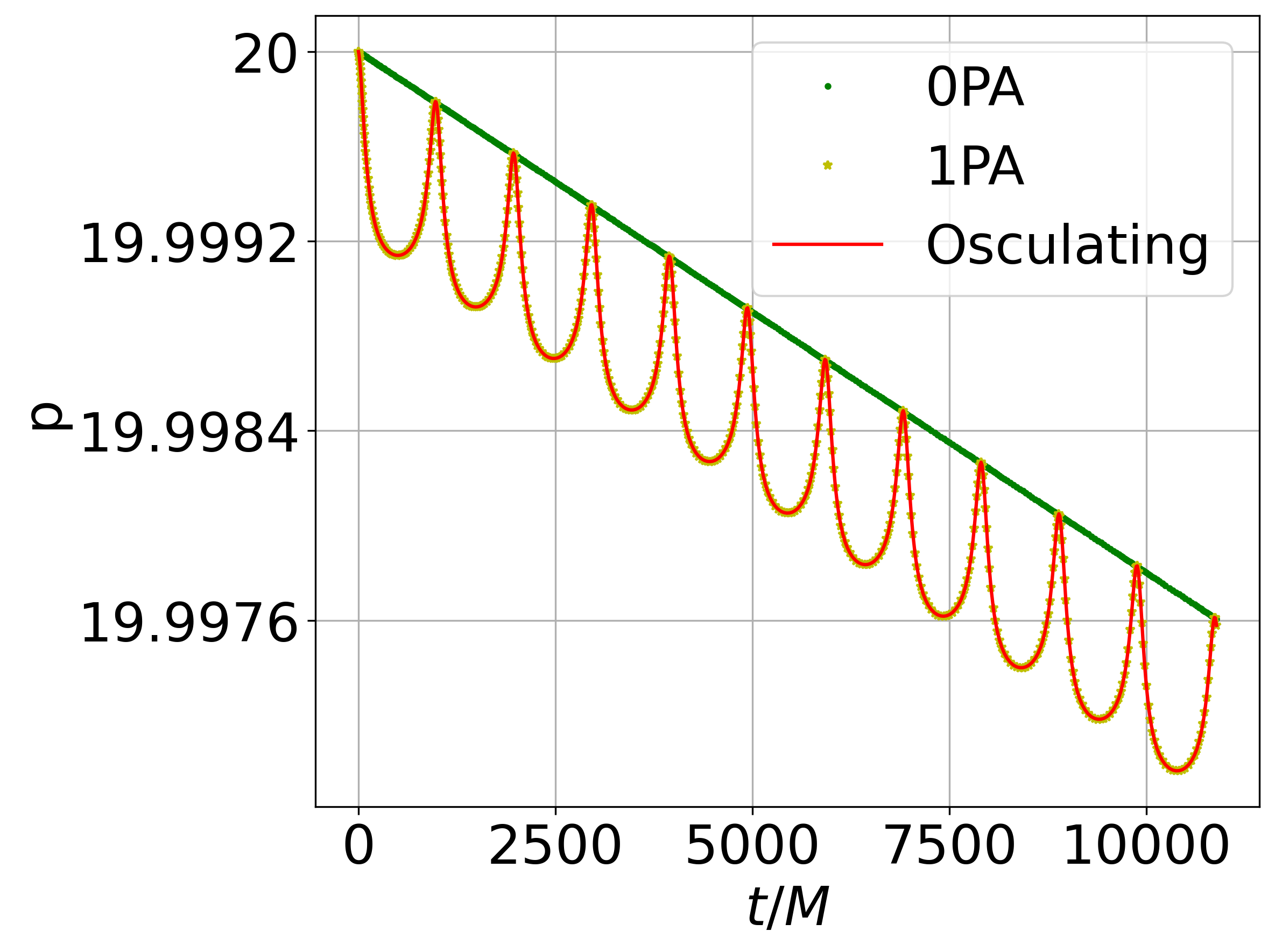}
    \end{subfigure}
    \hfill
    \begin{subfigure}[b]{0.32\textwidth}
        \includegraphics[width=\textwidth]{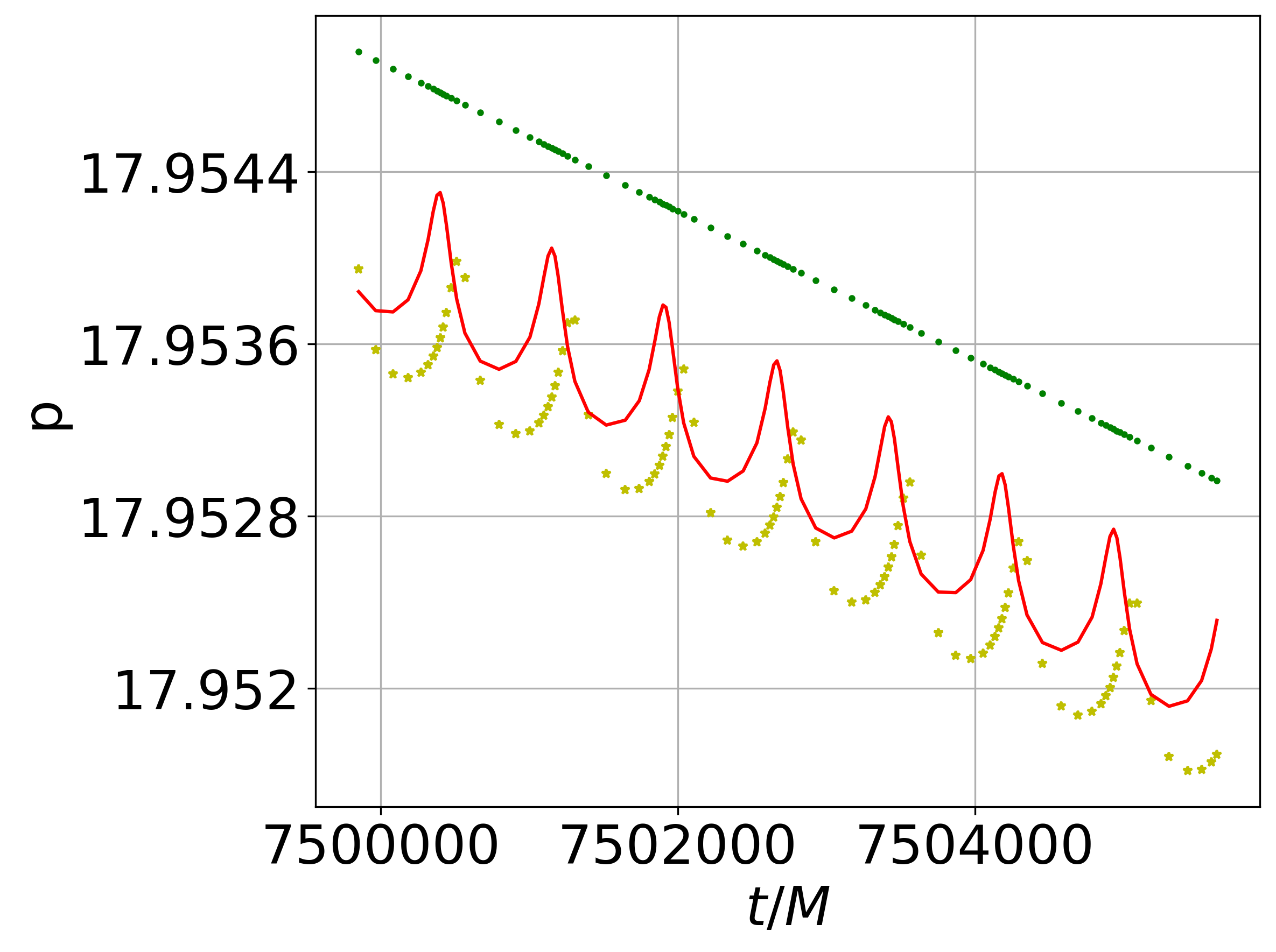}
    \end{subfigure}
    \hfill
    \begin{subfigure}[b]{0.32\textwidth}
        \includegraphics[width=\textwidth]{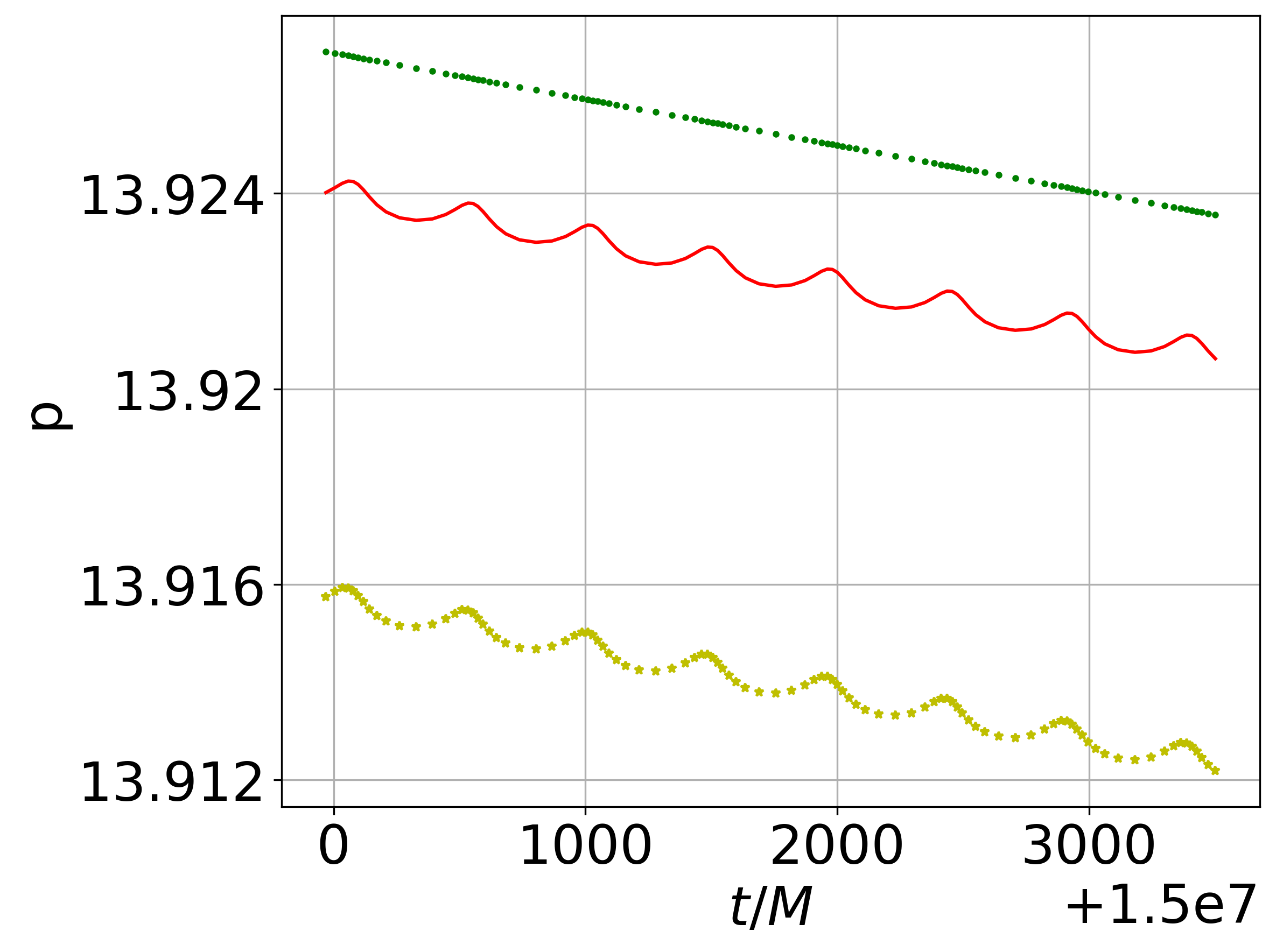}
    \end{subfigure}

    \vspace{1em}
    \begin{subfigure}[b]{0.32\textwidth}
        \includegraphics[width=\textwidth]{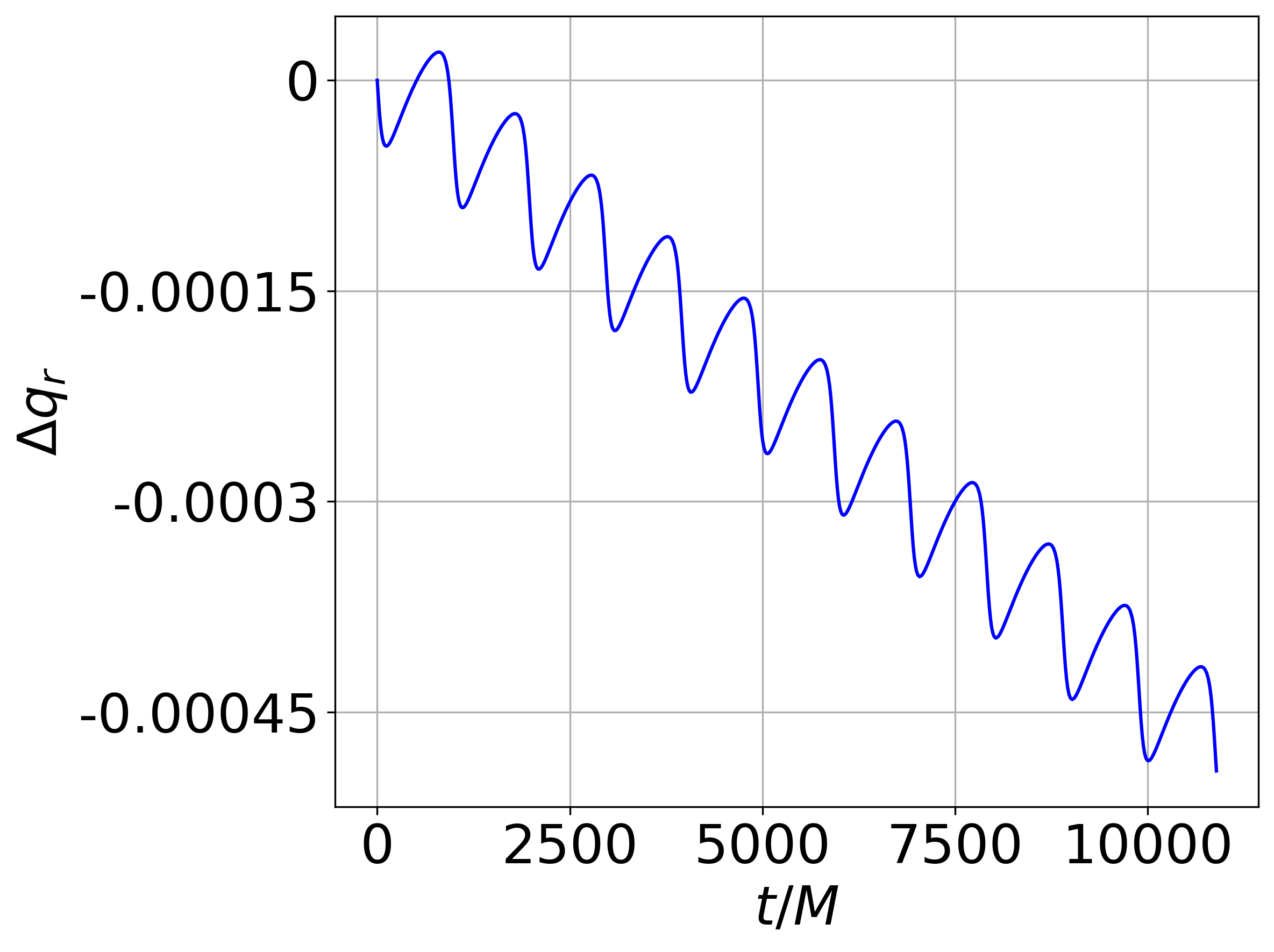}
    \end{subfigure}
    \hfill
    \begin{subfigure}[b]{0.32\textwidth}
        \includegraphics[width=\textwidth]{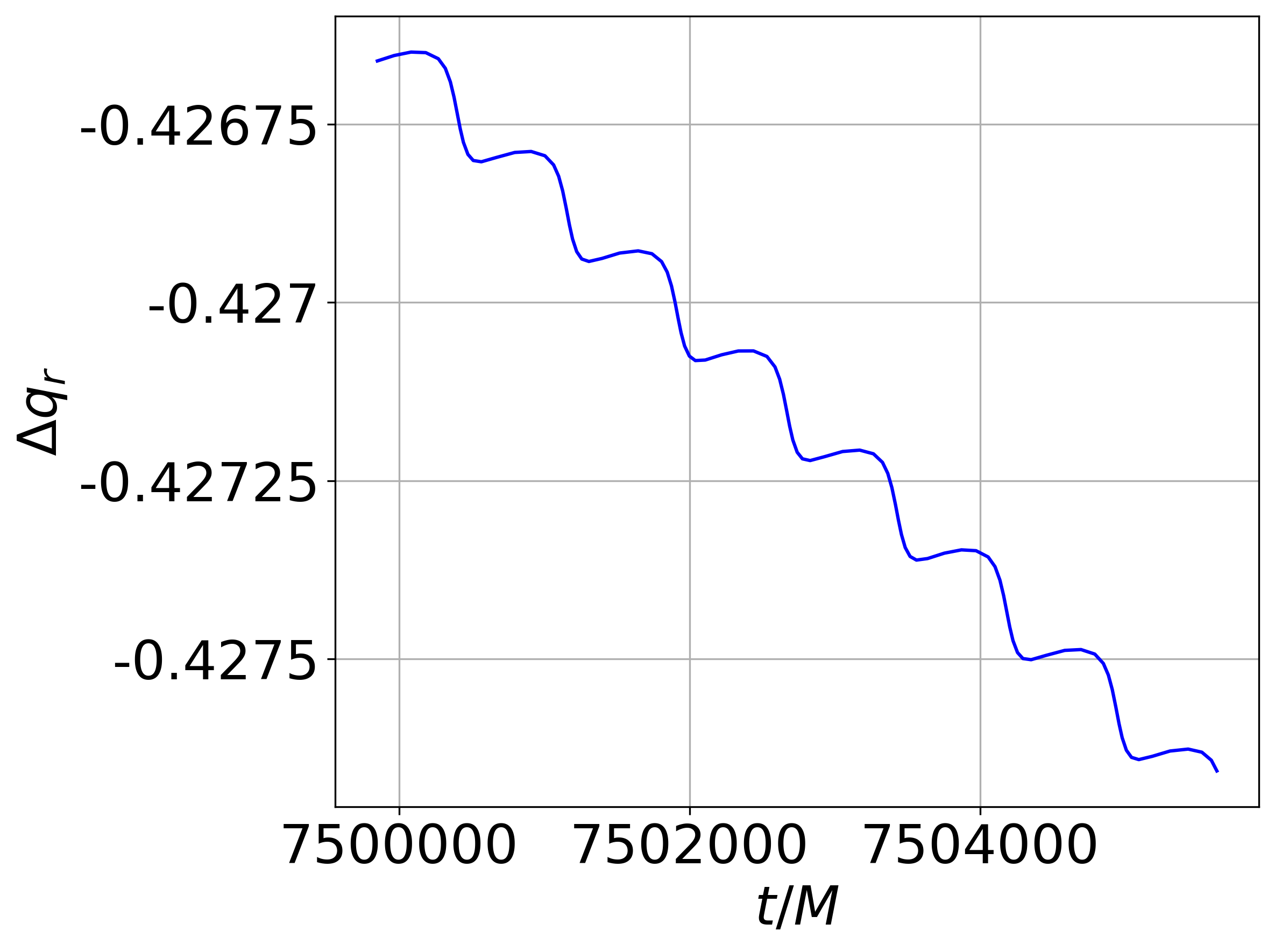}
    \end{subfigure}
    \hfill
    \begin{subfigure}[b]{0.32\textwidth}
        \includegraphics[width=\textwidth]{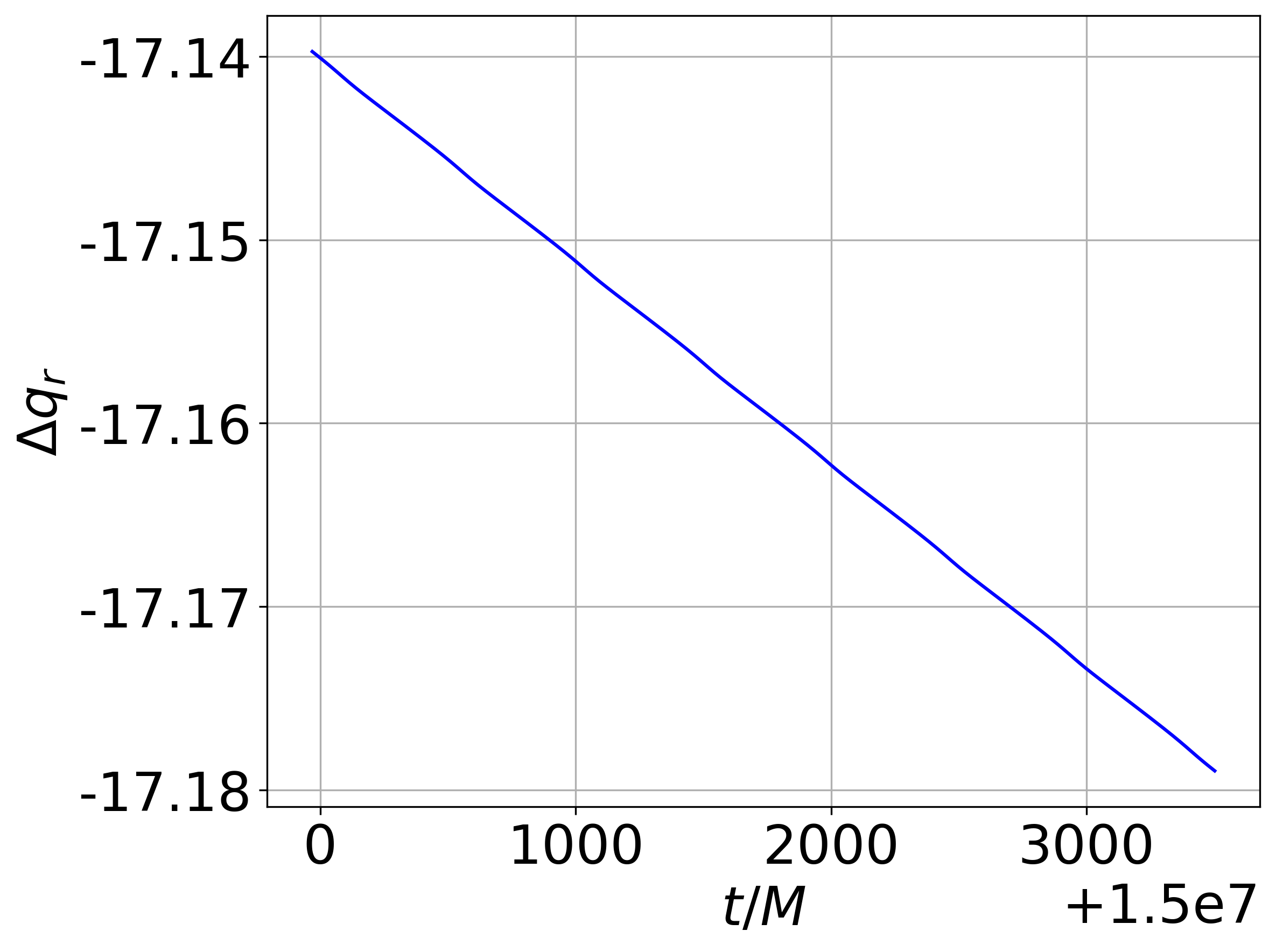}
    \end{subfigure}

    \vspace{1em}
    \begin{subfigure}[b]{0.32\textwidth}
        \includegraphics[width=\textwidth]{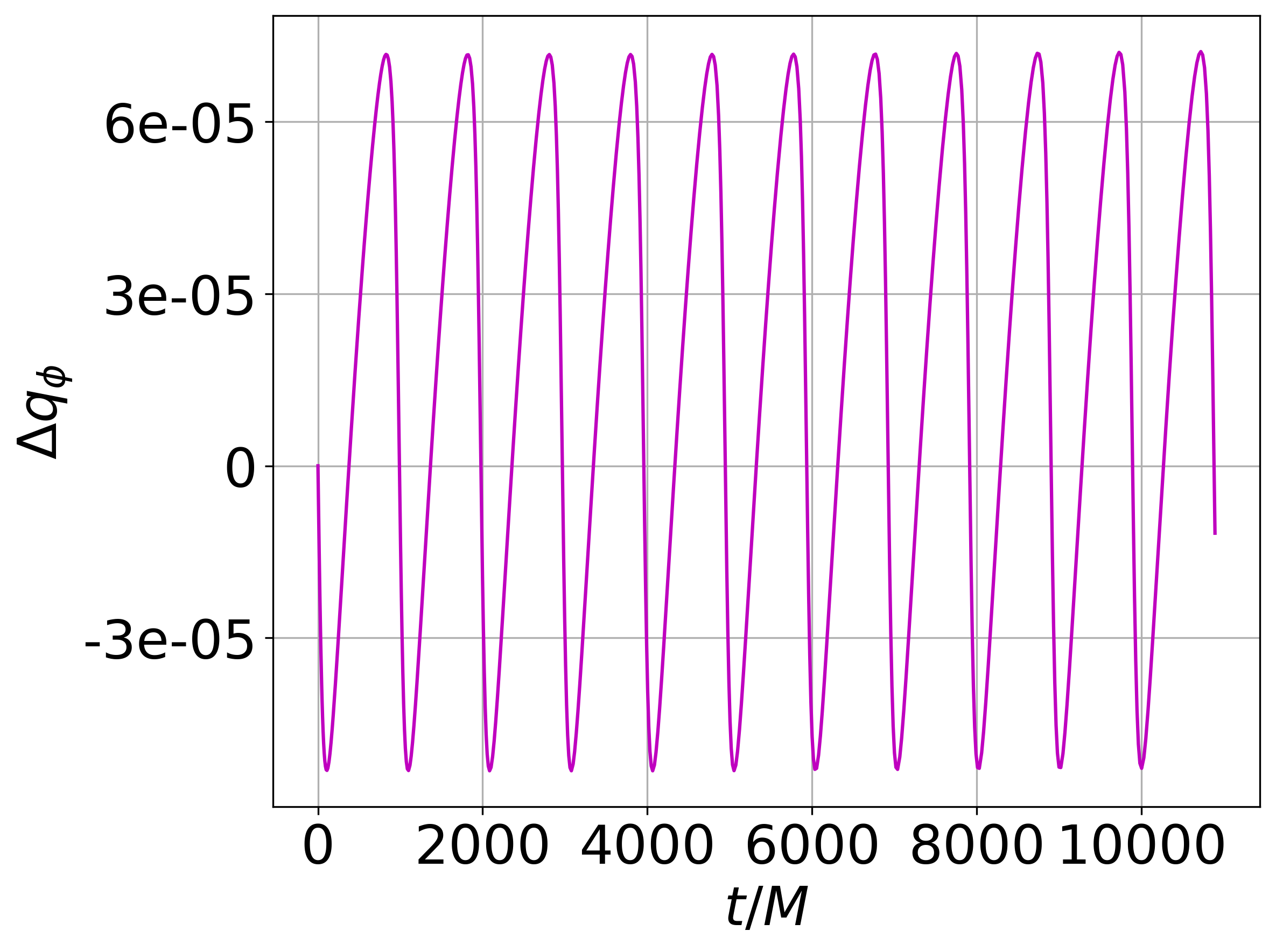}
    \end{subfigure}
    \hfill
    \begin{subfigure}[b]{0.32\textwidth}
        \includegraphics[width=\textwidth]{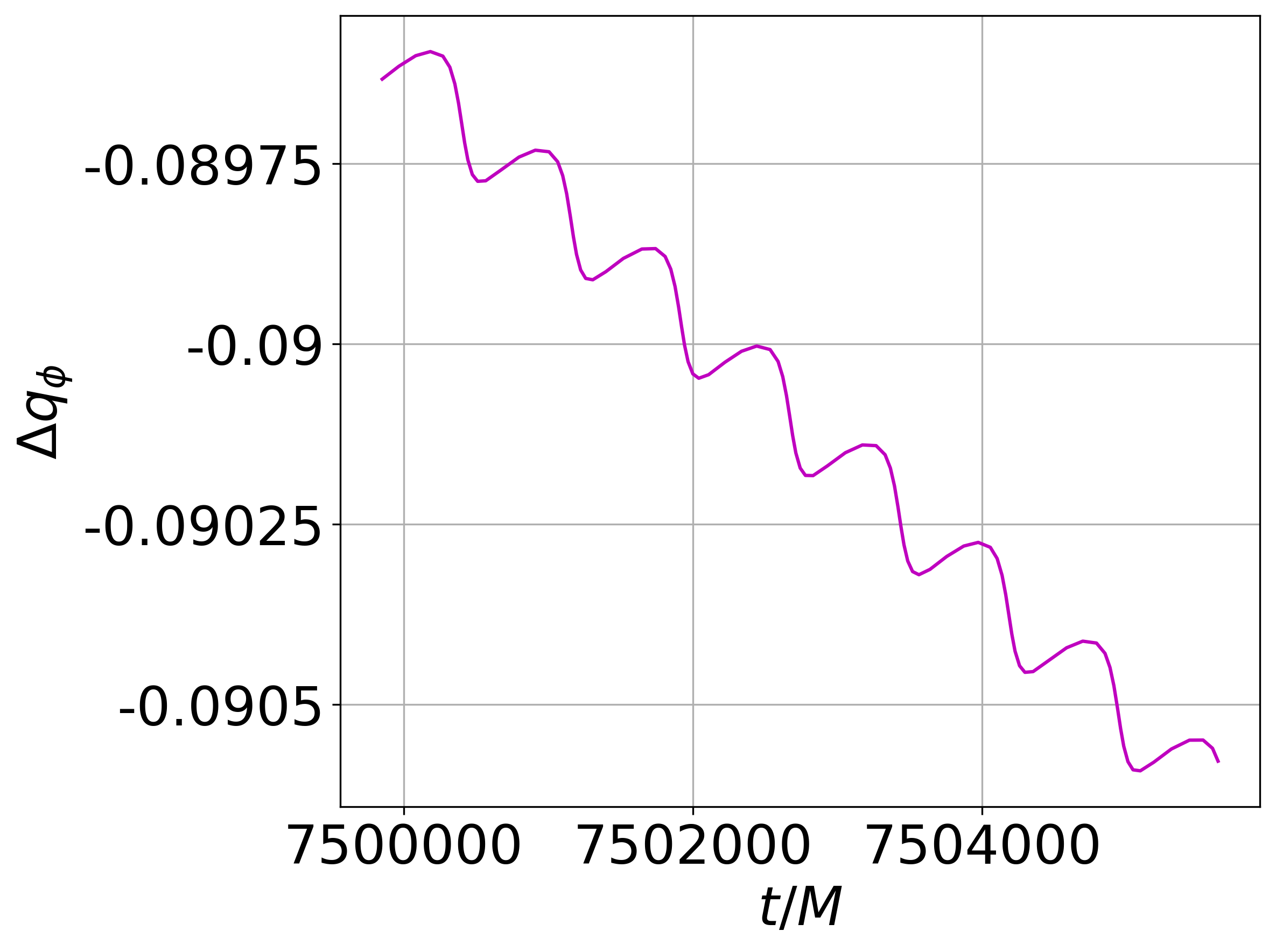}
    \end{subfigure}
    \hfill
    \begin{subfigure}[b]{0.32\textwidth}
        \includegraphics[width=\textwidth]{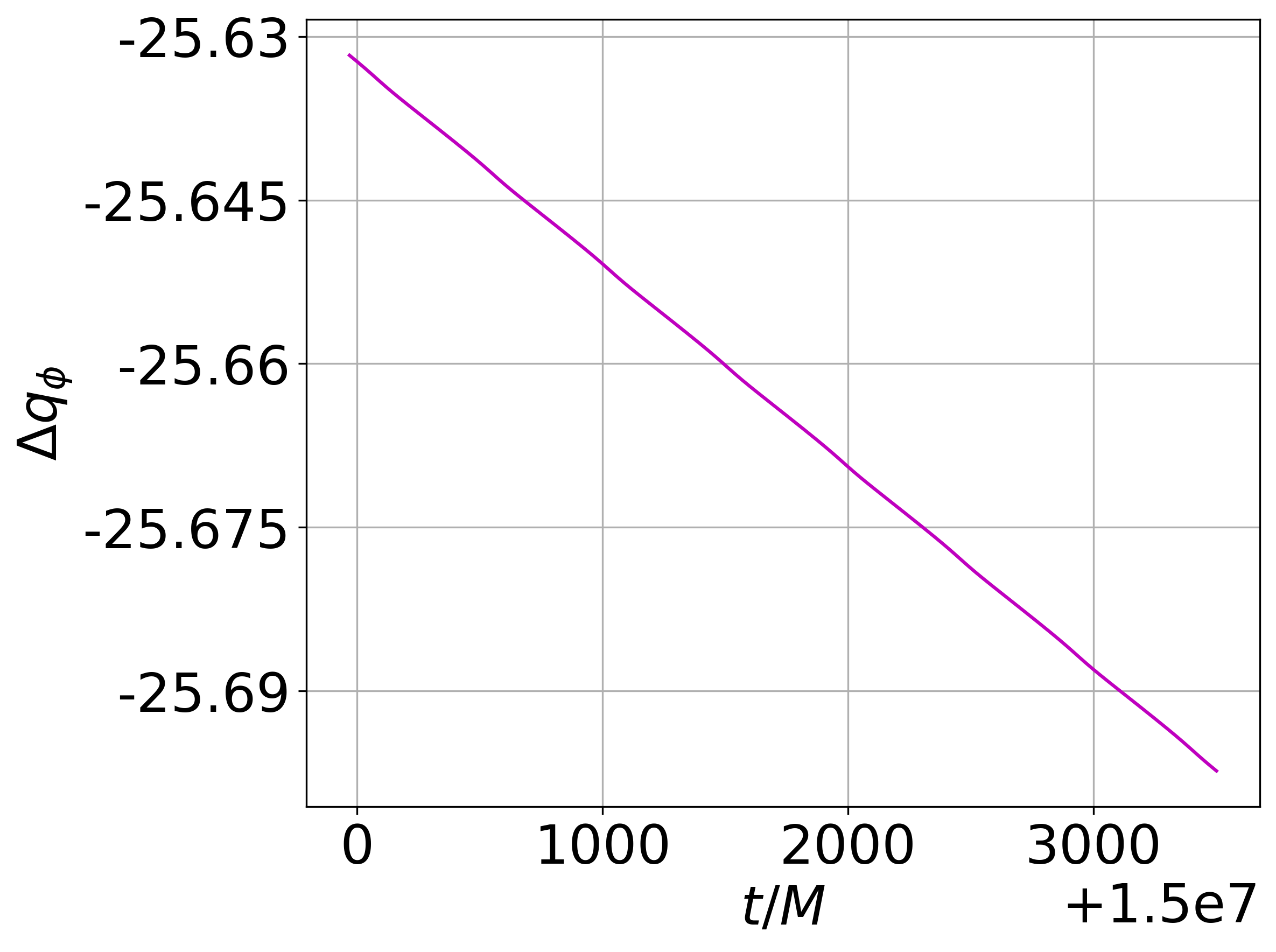}
    \end{subfigure}

    \caption{\raggedright The evolution of $(p,\Delta q_r,\Delta q_{\phi})$ as functions of dimensionless time for \acp{EMRI} with $\varepsilon=10^{-4},e=0.5$ at the initial (left column, $t\sim0$), middle (middle column, $t\sim7.5\times10^6M/\varepsilon$) , and final(right column,$t\sim1.5\times10^7M/\varepsilon$) evolutionary stages.}
    \label{fig:details}
\end{figure*}

For the $e=0.5$ case, we extracted three representative segments from the initial, middle, and final evolutionary stages, presenting their orbital parameter evolution details and phase errors in Fig.~\ref{fig:details}. The phase error exhibits slow linear growth throughout most of the evolution, maintaining a small magnitude of $\sim 10^{-2}$ over a remarkably long timescale  $\sim M/ \varepsilon$. The rather large phase error in the final segment arises from the system entering the plunge phase, where the two-timescale expansion assumptions break down.

\begin{figure}[htbp]
    \centering
    \includegraphics[width=0.9\linewidth]{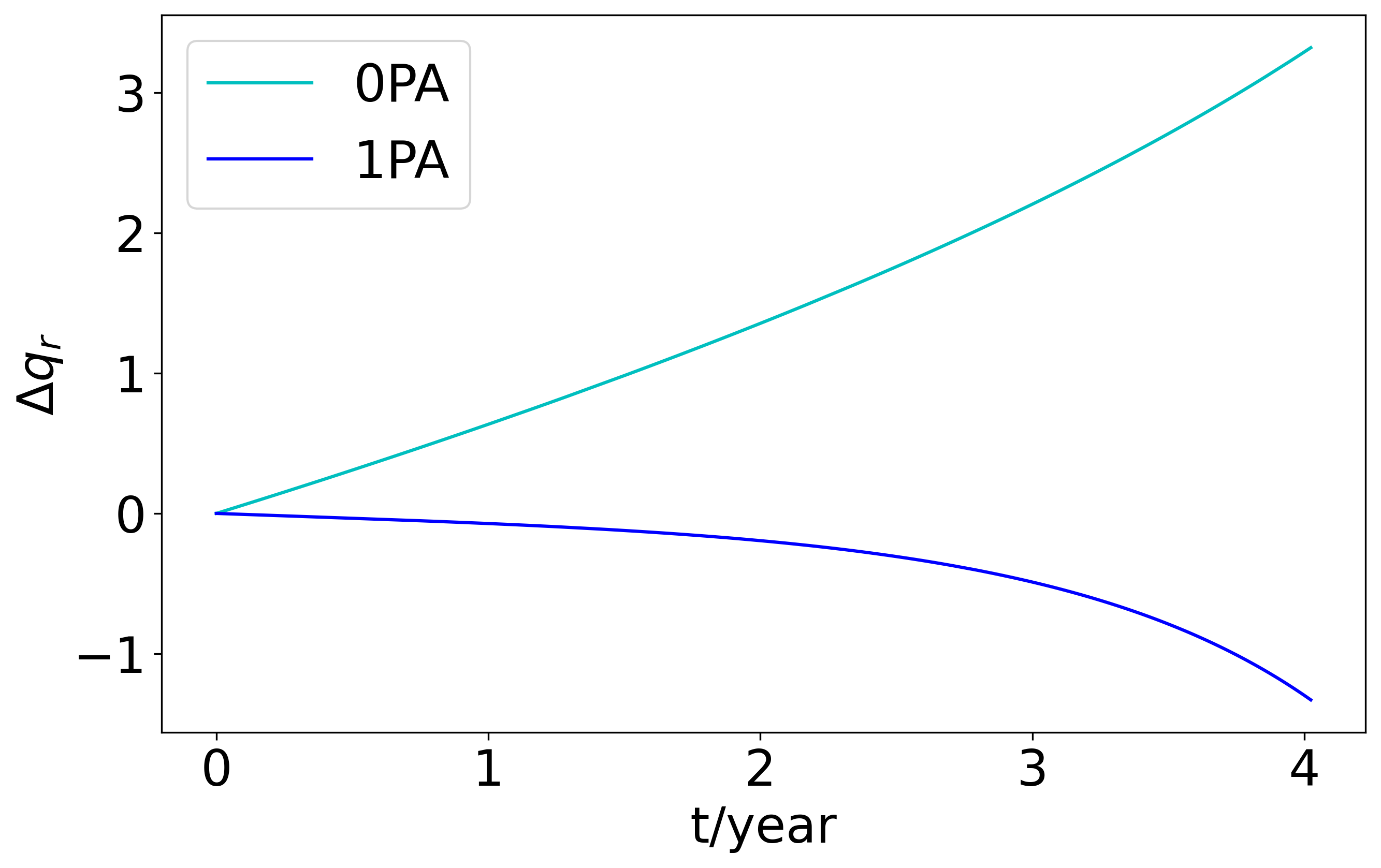}

    \includegraphics[width=0.9\linewidth]{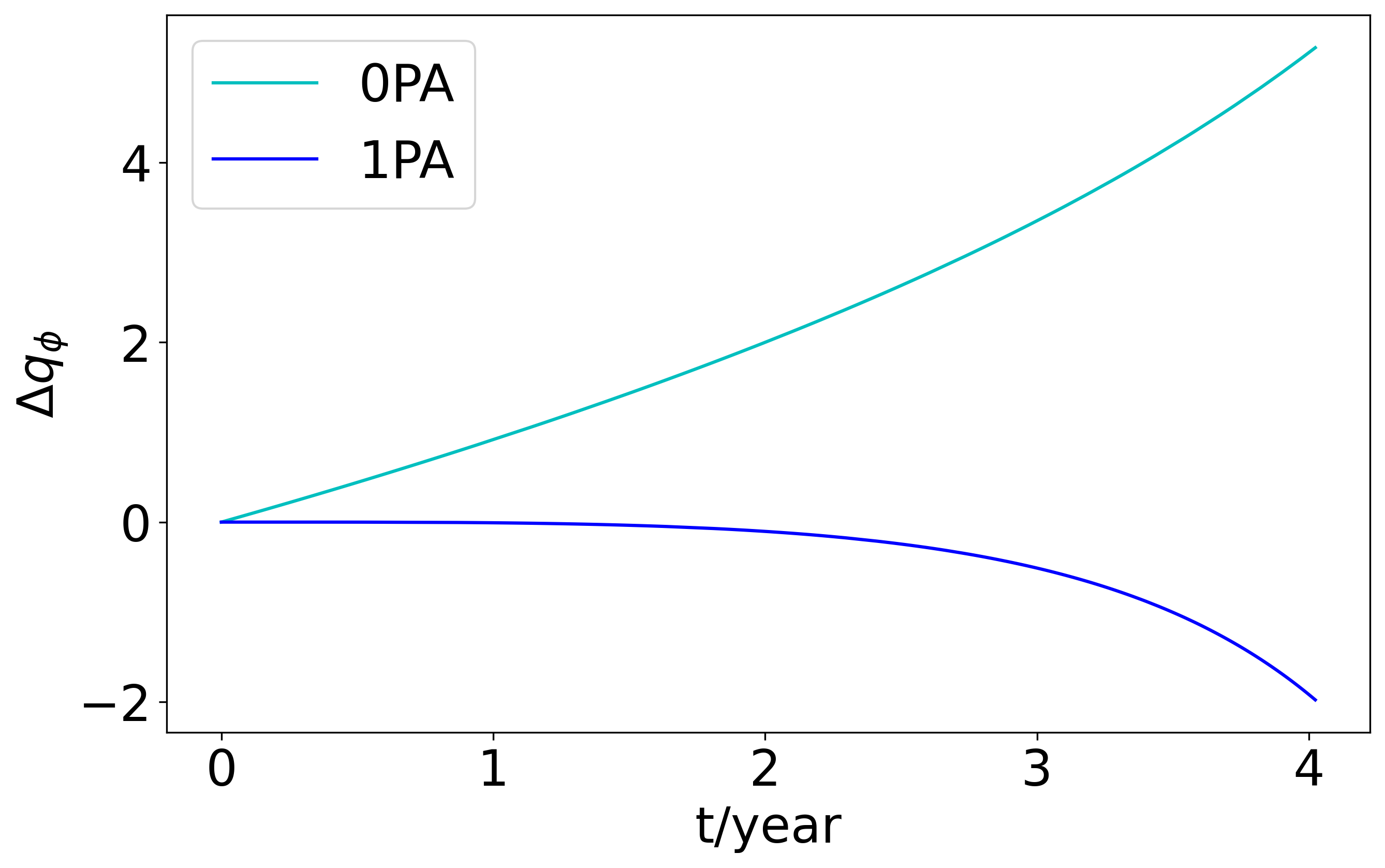}

    \caption{\raggedright The evolution of $(\Delta q_r,\Delta q_{\phi})$ as functions of time for \acp{EMRI} with  $(M,\varepsilon)=(10^6 M_{\odot},10^{-5})$ and begin the evolution at $(p,e)=(15,0.3)$. The blow up at the final stage is due to the break down of the two timescale expansion as it approaches to plunge.}
    \label{fig:years}
\end{figure}

In Fig.~\ref{fig:years}, we also compared the osculating geodesic method and the two-timescale expansion method for an EMRI system with parameters $(M,\varepsilon)=(10^6 M_{\odot},10^{-5})$ and the initial value for the orbit is chose to be $(p,e)=(15,0.3)$. The plunge happens at about 6 years after that, and the phase error blows up at the final stage. We plotted the dephasing for the first 4 years. The dephasing error remains less than 1 radian within the first 3.5-years, and less than 0.1 radian within the first 2 years.

We also calculated the orbital evolutions for $\varepsilon$ equals
to $10^{-3}$ and $10^{-4}$ for different initial eccentricities,
and the two timescale for 1PA always have a very good agreement with the osculating geodesic method.
We listed the computation cost for different methods in Table~\ref{tab:time}. The two timescale method (both 0PA and 1PA) achieves a reduction in computation time for several orders of magnitude compared to the osculating geodesic integration,
especially for small mass ratio cases.
As the mass ratio becomes smaller and smaller, the direct integration will cost
longer time if we want to keep the accuracy of the result.
However, the computation cost will not vary too much for two timescale calculations.
It should be noted that the computation time for different $\varepsilon$ are not exactly the same, so we only present the first significant digit in the table.

\begin{table}[htbp]
\centering
\caption{\raggedright The computation cost of the orbit evolution  for \acp{EMRI} with different values of initial eccentricity $e$ and mass ratio $\varepsilon$ with osculating geodesic (OG)method, and the two timescale expansion method for 0PA and 1PA.
The initial value of semi-latus rectum $p$ is $20M$.}
\label{tab:time}
\renewcommand{\arraystretch}{1.2}
\begin{tabular}{|c|c||c||c||c||}
\hline
$e$& $\varepsilon$ & OG & 0PA & 1PA \\
\hline \hline
\multirow{3}{*}{$0.1$}& $10^{-3}$ & $4$min &   &   \\
\cline{2-3}
  &$10^{-4}$ & $25$min & $10$s & $30$s \\
  \cline{2-3}
  & $10^{-5}$ & $168$min &   &   \\
  \hline \hline
  \multirow{3}{*}{$0.5$}& $10^{-3}$ & $4$min &   &   \\
  \cline{2-3}
  &$10^{-4}$ & $25$min & $20$s & $50$s \\
  \cline{2-3}
  &$10^{-5}$ & $139$min &   &   \\
  \hline \hline
  \multirow{3}{*}{$0.9$}& $10^{-3}$ & $3$min &   &   \\
\cline{2-3}
  &$10^{-4}$ & $25$min & $30$s & $90$s \\
  \cline{2-3}
  &$10^{-5}$ & $133$min &   &  \\
  \hline
\end{tabular}
\end{table}

On the other hand, a near identity transformation (NIT) method was employed
in \cite{VanDeMeent:2018cgn} to derive the evolution equations on the Schwarzschild background.
Their approach using NIT, also gives the same results with the osculating geodesic method, and thus it should give the same results
with the two timescale expansion method at the same order.
However,the NIT equations still contain $\varepsilon$, which must be specified before solving equations.
But the two timescale equations presented here, on the other hand,
can be solved for all possible values of $\varepsilon$ in the same step.
Given our goal of generating the waveform templates rapidly for a wide range of parameters,
this method offers significant advantages in reducing the cost of numerical computations.

\section{Conclusions}\label{con}

In this paper, we have systematically established the theoretical framework
to calculate the \ac{EMRI} waveform for eccentric orbits on Schwarzschild background.
We derived the expression of puncture field and deduced the equation of motions
into a series ODEs in frequency domain via two timescale method.
We also presented the numerical calculation of the orbital evolution with osculating geodesic method and two timescale method of 0PA and 1PA.
We find that the 1PA two timescale results have a very good agreement with the osculating geodesic method.
These results establish the necessary analytical equations for
generating \ac{EMRI} waveform templates on Schwarzschild background,
making a step toward the final goal of general \ac{EMRI} waveform templates on Kerr background
which corresponding to realistic systems.

In the future, based on the current results,
we will focus on developing numerical methods for rapidly generating generic Schwarzschild templates,
to address potential computational challenges in waveform modeling.

The way to the most general EMRI waveform templates is sketched in \cite{Pound:2021qin}.
However, due to the complexity of the Kerr spacetime,
some new obstacles emerge when we introduce the spin of central blackhole,
such as the orbital resonances\cite{Speri:2021psr} and
the reconstruction of metric perturbations for 2nd order \ac{SF} calculation\cite{Whiting:2005hr}. Although there are encouraging progress in overcoming these obstacles
\cite{Lynch:2024ohd,Berens:2024czo,Hollands:2024iqp},
A practical waveform template algorithm for data analysis remains a long-term goal.
A simpler test model for generic Schwarzschild EMRI is therefore an important intermediate goal,
which is the main progress in this paper.

\begin{acknowledgments}

This work is supported by the National Key Research and~Development Program of China (Grant No. 2023YFC2206703, 2021YFC2203002),  the Guangdong Basic and Applied Basic Research Foundation(Grant No. 2023A1515030116), and the National Science Foundation of China (Grant No. 12261131504).
Many calculations in this paper rely on xAct\cite{xAct,Martin-Garcia:2008ysv,Brizuela:2008ra}.

\end{acknowledgments}

\appendix

\section{The Defination of Elliptic Integrals}\label{app:ell}

The parameters $(\alpha, \beta, \gamma)$ are defined as $\left( \frac{2 x}{1 + x}, \frac{2 y}{1 - y},
\frac{2 z}{1 - z} \right)$.
The elliptic integrals are defined as

\begin{eqnarray*}
  F (\phi | \beta) & = & \int^{\phi}_0 \frac{\mathd \theta}{\sqrt{1
  - \beta \sin^2 \theta}};\\
  E (\phi | \beta) & = & \int^{\phi}_0 \mathd \theta \sqrt{1 -
  \beta \sin^2 \theta};\\
  \Pi (\alpha ; \phi | \beta) & = & \int^{\phi}_0 \frac{\mathd
  \theta}{(1 - \alpha \sin^2 \theta) \sqrt{1 - \beta \sin^2 \theta}};
\end{eqnarray*}
where $F (\phi | \beta),E (\phi | \beta),\Pi (\alpha ; \phi | \beta)$
is incomplete elliptic integral of the first, second, third kind, respectively.
While
\begin{eqnarray*}
  K (\beta) & = &F (\dfrac{\pi} {2} | \beta) ,\\
  E (\beta) & = &E (\dfrac{\pi} {2} | \beta),\\
  \Pi (\alpha | \beta) & = & \Pi \left( \alpha ; \frac{\pi}{2}
  | \beta \right).
\end{eqnarray*}
are the corresponding complete elliptic integrals.
Since it's difficult to directly solve Eq.\eqref{Ellp}
with current symbolic computation software,
we need some elliptic integral identities in \cite{Byrd:1971bey} to simplify the results.
Note that the definitions in \cite{Byrd:1971bey} are slightly different with us.

The function $I_2 (x, y, z ; \chi)$ is a composition of elliptic integrals defined as
\begin{widetext}

\begin{eqnarray*}
  I_2 (x, y, z ; \chi)& = & \frac{2}{(1 + x)^2 (1 - y) \sqrt{1 - z}} \times\left[\frac{\alpha^2}{2 (\alpha - 1) (\alpha - \beta) (\gamma - \alpha)} E
  \left( \frac{\chi}{2} \Big| \gamma \right) + \frac{\alpha}{2 (\alpha - 1)
  (\alpha - \beta)} F \left( \frac{\chi}{2} \Big| \gamma \right)\right.\\
  &+& \left( \frac{\alpha (2 \alpha \gamma + 2 \alpha - \alpha^2 - 3 \gamma)}{2
  (\alpha - 1) (\alpha - \beta) (\gamma - \alpha)}-\frac{\alpha \beta}{(\alpha - \beta)^2}\right) \Pi \left( \alpha ;
  \frac{\chi}{2} \Big| \gamma \right) +\frac{\beta^2}{(\alpha -\beta)^2} \Pi \left(\beta;\frac{\chi}{2} \Big| \gamma \right)\\
  &-& \left.
  \frac{\alpha^3 \sin \chi}{2 \sqrt{2} (\alpha - 1) (\alpha - \beta)
  (\gamma - \alpha)} \frac{ \sqrt{2 - \gamma + \gamma \cos \chi}}{ 2 - \alpha
  + \alpha \cos \chi}\right]
\end{eqnarray*}

\end{widetext}

\section{Coefficients of Puncture Field}\label{app:pun}
 Here we give the explicit expressions of non-zero coefficients appear in Eq. \eqref{punc}.

\begin{flalign*}
  &A_{00} = \frac{4(r-2M)^2}{r^2}(u^t)^2, &&A_{01} = -4u^tu^r,&\\
  &A_{03} = -4r(r-2M)u^tu^\phi, &&A_{11} = \frac{4r^2}{(r-2M)^2}(u^r)^2,&\\
  &A_{13} = \frac{4r^3}{r-2M}u^ru^\phi, &&A_{33} = 4r^4 (u^\phi)^2.&
\end{flalign*}

\begin{flalign*}
  &B_{00}^r = \frac{2M(r-2M)}{r^3}(u^t)^2, &&B_{01}^r = \frac{6M}{r(r-2M)}u^tu^r,&\\
  &B_{01}^\phi = -4(r-2M)u^tu^\phi, &&B^{\theta}_{02} = 4(r-2M)u^tu^r,&\\
  &B_{03}^r = -2(2r-5M)u^tu^\phi, &&B_{03}^\phi = 2(r-2M)u^tu^r,&\\
  &B_{11}^r = -\frac{14Mr}{(r-2M)^3}(u^r)^2,
  &&B_{11}^{\phi} = \frac{8r^2}{r-2 M}u^ru^\phi,&\\
  &B^{\theta}_{12} = \frac{4r^2}{r-2M}(u^r)^2,
  &&B^r_{13} = \frac{2r^2(2r-9M)}{(r-2M)^2}u^ru^\phi,&\\
  &B^{\phi}_{13} =4r^3(u^\phi)^2-\frac{2r^2}{r-2M}(u^r)^2,
  &&B^{\theta}_{23} = 4r^3 u^r u^\phi,&\\
  &B^r_{33} = \frac{2r^3(4r-11M)}{r-2M} (u^\phi)^2, &&B^\phi_{33} = - 6r^3u^ru^\phi.
\end{flalign*}

\begin{eqnarray*}
     C^{rr\phi}_{00} & = & 2 (2r-M) (u^t)^2 u^r u^{\phi}, \\
     C^{r\theta\theta}_{00} & =&  \frac{2(r-2M)}{r} [3r-3M+r(u^r)^2](u^t)^2 ,\\
     C^{r\phi\phi}_{00} & =&  \frac{2(r-2M)}{r} [3r-3M+r(u^r)^2+r^2(2r-M)(u^\phi)^2](u^t)^2, \\
     C^{\theta\theta\phi}_{00} & =&  C^{\phi\phi\phi}_{00} = 2  r (r - 2 M)^2(u^t)^2 u^r u^\phi ,\\
     C^{ijk}_{01}&=&-\frac{r^2}{(r-2M)^2}\frac{u^r}{u^t}C^{ijk}_{00},~~~
     C^{ijk}_{11}=\frac{r^4}{(r-2M)^4}\frac{(u^r)^2}{(u^t)^2}C^{ijk}_{00}.
\end{eqnarray*}

\begin{eqnarray*}
     C^{rr\phi}_{03} & =&  \frac{2r}{r-2M}
     [r-2M+r(u^r)^2-r^2(2r-M)(u^\phi)^2]u^tu^r,\\
     C^{r\theta\theta}_{03} & = &-2r^2[3r-3M+r(u^r)^2]u^tu^\phi,\\
     C^{r\phi\phi}_{03} & =& -2r^2[3r-3M-r(u^r)^2+r^2(2r-M)(u^\phi)^2]u^tu^\phi,\\
     C^{\theta\theta\phi}_{03} & =& 2r^2(r-2M)[1-r^2(u^\phi)^2]u^tu^r,\\
     C^{\phi\phi\phi}_{03} & =& 2r^2(r-2M)u^tu^r,\\
     C^{ijk}_{13}&=&-\frac{r^2}{(r-2M)^2}\frac{u^r}{u^t}C^{ijk}_{03},~~~
     C^{ijk}_{33}=-\frac{r^3}{r-2M}\frac{u^\phi}{u^t}C^{ijk}_{03}.
\end{eqnarray*}

\section{Explicit Expressions for Two-Timescale Expansion Terms}\label{app:tte}
Here we give the explicit expressions in  Eq.\eqref{qq} and Eq.\eqref{II}.
Only $G_a^{(1)},G_a^{(2)}$ and $g_r^{(1)}$ are needed for a 1PA calculation.
$\chi$ appears here should be regarded as $\chi(q_r^{(0)})$.

\begin{widetext}

\begin{flalign}
  G^{(1)/(2)}_p&=f^{(1)/(2)}_r
  \times\sqrt{\frac{p_0-6-2e_0\cos\chi}{(p_0-2)^2-4e_0^2}}
  \times\frac{2p_0(p_0-3-e_0^2)}{(p_0-6)^2-4e_0^2}
  \times[(2-p_0)e_0\sin\chi+e_0^2\sin2\chi]&\nonumber\\
  &-f^{(1)/(2)}_{\phi}\times\sqrt{\frac{p_0}{(p_0-2)^2-4 e_0^2}}
  \times\frac{Mp_0(p_0-3-e_0^2)}{[(p_0-6)^2-4e_0^2](1+e_0\cos\chi)^2}&\nonumber\\
  &\hspace{29pt}\times\{~[3(24+8e_0^2+e_0^4)-12(6+e_0^2)p_0+(22+e_0^2)p_0^2-2p_0^3]
  +[24(4+e_0^2)-2(28+3e_0^2)p_0+8p_0^2]e_0\cos\chi&\nonumber\\
  &\hspace{37.5pt}+[4(6+e_0^2)-12p_0+p_0^2]e_0^2\cos2\chi+2(4-p_0)e_0^3\cos3\chi+e_0^4\cos4\chi\}&
\end{flalign}

\begin{flalign}
  G^{(1)/(2)}_e&=f^{(1)/(2)}_r\times\sqrt{\frac{p_0-6-2e_0\cos\chi}{(p_0-2)^2-4e_0^2}}
  \times\frac{(p_0-3-e_0^2)}{e_0[(p_0-6)^2-4e_0^2]}&\nonumber\\
  &\hspace{29pt}\times\{~(2+p_0)e_0+[4(3-e_0^2)-8p_0+p_0^2]\cos\chi
  +(6-p_0)e_0\cos2\chi\}&\nonumber\\
  &-f^{(1)/(2)}_{\phi}\times\sqrt{\frac{p_0}{(p_0-2)^2-4e_0^2}}
  \times\frac{M}{2e_0[(p_0-6)^2-4e_0^2](1+e_0\cos\chi)^2}&\nonumber\\
  &\hspace{29pt}\times\{~[4(108+72e_0^2+9e_0^4-e_0^6)-4(144+63e_0^2+4e_0^4)p_0
  +2(138+35e_0^2+e_0^4)p_0^2-2(28+3e_0^2)p_0^3+4p_0^4]\sin\chi&\nonumber\\
  &\hspace{37.5pt}+[4(108+39e_0^2+e_0^4)-2(216+47e_0^2-e_0^4)p_0
  +2(75+8e_0^2)p_0^2-(21+e_0^2)p_0^3+p_0^4]e_0\sin2\chi&\nonumber\\
  &\hspace{37.5pt}+[4(36+9e_0^2-e_0^4)-4(27+4e_0^2)p_0
  +2(13+e_0^2)p_0^2-2p_0^3]e_0^2\sin3\chi&\nonumber\\
  &\hspace{37.5pt}+[6(3+e_0^2)-(9+e_0^2) p_0+p_0^2]e_0^3\sin4\chi\}&
\end{flalign}

\begin{flalign}
  g^{(1)}_r&=f^{(1)}_r
  \times\frac{M\Omega_rp_0^2(p_0-3-e_0^2)}{e_0[(p_0-6)^2-4e_0^2](1+e_0\cos\chi)^2}
  \times[2e_0-(6-p_0)\cos\chi]&\nonumber\\
  &+f^{(1)}_{\phi}\times\sqrt{\frac{p_0}{p_0-6-2e_0\cos\chi}}
  \times\frac{M^2\Omega_rp_0^2(p_0-3-e_0^2)}{2e_0[(p_0-6)^2-4e_0^2](1+e_0\cos\chi)^4}&\nonumber\\
  &\hspace{29pt}\times\{~[6(12+e_0^2)-(36+e_0^2)p_0+4p_0^2]\sin\chi
  +[4(9-e_0^2)-12p_0+p_0^2]e_0\sin2\chi+(6-p_0)e_0^2\sin3\chi\}&
\end{flalign}
\end{widetext}

\bibliographystyle{unsrtnat}
\bibliography{Ecc2SF}

\end{document}